\documentclass[pdflatex,sn-nature]{sn-jnl}
\usepackage{graphicx}%
\usepackage{multirow}%
\usepackage{amsmath,amssymb,amsfonts}%
\usepackage{amsthm}%
\usepackage{mathrsfs}%
\usepackage[title]{appendix}%
\usepackage{xcolor}%
\usepackage{textcomp}%
\usepackage{manyfoot}%
\usepackage{booktabs}%
\usepackage{algorithm}%
\usepackage{algorithmicx}%
\usepackage{algpseudocode}%
\usepackage{listings}%
\usepackage{caption}
\usepackage{natbib}
\usepackage{dcolumn}
\usepackage{siunitx}
\DeclareSIUnit{\gauss}{G}
\DeclareRobustCommand{\hyphenSI}[2]{\SI[number-unit-product={\text{-}}]{#1}{#2}}
\usepackage{lineno}
\usepackage{braket}
\setcitestyle{super}
\graphicspath{{Figures/Fig}{images/}}

\begin{document}
\title[Article Title]{Heterogeneous entanglement between a trapped ion and a solid-state quantum memory}

\author[1,2,3]{\fnm{Chen-Xu} \sur{Wang}}
\equalcont{These authors contributed equally to this work.}

\author[1,2,3]{\fnm{Yi-Yang} \sur{Wang}}
\equalcont{These authors contributed equally to this work.}

\author[1,2,3]{\fnm{Tian-Xiang} \sur{Zhu}}
\equalcont{These authors contributed equally to this work.}

\author[1,2,3]{\fnm{Qing-Quan} \sur{Yao}}
\equalcont{These authors contributed equally to this work.}

\author[1,2,3]{\fnm{Peng-Jun} \sur{Liang}}

\author[1,2,3]{\fnm{Yuan-Cong} \sur{Li}}

\author[1,2,3]{\fnm{Zi-Peng} \sur{Liu}}

\author[5]{\fnm{Ran} \sur{He}}

\author[1,2,3,4]{\fnm{Yong-Jian} \sur{Han}}

\author*[1,2,3,4]{\fnm{Jin-Ming} \sur{Cui}}\email{jmcui@ustc.edu.cn}

\author*[1,2,3,4]{\fnm{Zong-Quan} \sur{Zhou}}\email{zq\_zhou@ustc.edu.cn}

\author*[1,2,3,4]{\fnm{Yun-Feng} \sur{Huang}}\email{hyf@ustc.edu.cn}

\author*[1,2,3,4]{\fnm{Chuan-Feng} \sur{Li}}\email{cfli@ustc.edu.cn}

\author[1,2,3,4]{\fnm{Guang-Can} \sur{Guo}}

\affil[1]{\orgdiv{Laboratory of Quantum Information}, \orgname{University of Science and Technology of China}, \orgaddress{\city{Hefei}, \postcode{230026}, \country{China}}}

\affil[2]{\orgdiv{Anhui Province Key Laboratory of Quantum Network}, \orgname{University of Science and Technology of China}, \orgaddress{\city{Hefei}, \postcode{230026}, \country{China}}}

\affil[3]{\orgdiv{CAS Center For Excellence in Quantum Information and Quantum Physics}, \orgname{University of Science and Technology of China}, \orgaddress{\city{Hefei}, \postcode{230026}, \country{China}}}

\affil[4]{\orgdiv{Hefei National Laboratory}, \orgname{University of Science and Technology of China}, \orgaddress{\city{Hefei}, \postcode{230088}, \country{China}}}

\affil[5]{\orgname{Unitary Quantum Co., Ltd.}, \orgaddress{\city{Hefei}, \country{China}}}

\abstract{
Hybrid quantum networks offer a promising architecture for scalable quantum information processing and a future quantum internet, as they can combine the complementary strengths of disparate physical platforms. While single-atom systems provide deterministic quantum logic gates, atomic ensembles enable large-capacity quantum storage. However, generating entanglement between such heterogeneous systems has remained an open challenge, primarily due to fundamental spectral
mismatches and system complexity.
Here, we demonstrate a hybrid quantum network that entangles a single trapped $\mathrm{^{171}Yb^{+}}$ ion and a quantum memory based on $\rm ^{153}Eu^{3+}\colon\!Y_2SiO_5$ crystal over a \hyphenSI{75}{\meter} separation. Using polarization-maintaining quantum frequency conversion, we map spin-photon entanglement onto a hybrid entanglement between a single spin qubit and a collective excitation of the quantum memory. The resulting entangled state achieves a fidelity of $(89.2\pm2.3)\%$ and violates the Bell inequality by 6 standard deviations ($S = 2.328\pm0.054$) between two heterogeneous nodes. This work establishes entanglement between a quantum processing module with an ensemble-based quantum memory node, representing a key step toward a scalable, multifunctional quantum internet.
}



\maketitle

A quantum internet~\cite{Kimble2008,Wehner:2018enf} would enable secure quantum communication, distributed quantum computing, and distributed quantum sensing by interconnecting diverse quantum nodes.
Remote entanglement between atomic quantum nodes has been pursued for nearly two decades, with demonstrations using atomic ensembles~\cite{Chou2005Nat,Chou2007Sci,Liu2024}, trapped ions~\cite{Moehring2007,Krutyanskiy2023PRL,Main_2025,Liu2026Nat}, single atoms~\cite{Julian2012Sci,Ritter_2012,vanLeent2022,Lu2026Sci}, color centers~\cite{Bernien2013Nat,Hensen2015Nat,Knaut2024,Arian2024SciAdv}, semiconductor quantum dots~\cite{Delteil2016NatPhy,Stockill2017PRL}, and rare-earth-doped crystals~\cite{Lago-Rivera2021Nat,Liu_2021,Ruskuc2025Nat,Zhu2026natpho}. However, these demonstrations have predominantly connected nodes based on the same physical platform.  A notable achievement is the entanglement between a single atom and a Bose-Einstein condensate~\cite{Muecke2011PRL}, which connects two different physical platforms within the same atomic species. 
The growing complexity of target applications further requires hybrid quantum interconnects that leverage the complementary advantages of heterogeneous quantum systems~\cite{PRXQuantum.2.017002, Xiang2013RevModPhys,Azuma2023RevModPhys,David2025ci, Gu2025npj}.

Combining a quantum processor with individual atom control with a multiplexed, ensemble-based quantum memory can dramatically reduce the number of physical qubits required for practical tasks such as factoring RSA integers~\cite{PRL.127.140503}. The inherent multiplexing capability of such memories also promises a substantial increase in the entanglement distribution rate between distant quantum processors~\cite{tissot2025,Cussenot2025prl,liu2024NC}
, establishing a powerful hybrid architecture for large-scale quantum computing networks. Consequently, the generation of heterogeneous entanglement between a quantum processor and ensemble-based quantum memories represents a pivotal milestone for advancing next-generation quantum networks~\cite{Kimble2008,Wehner:2018enf,PRXQuantum.2.017002}.
Heterogeneous quantum interfaces have made important progress~\cite{Maring2017,PRL.123.213601,tang2015NC, sciadv.adi7346}, demonstrating the key compatibility for single-photon interfaces, but have not yet established a shared entanglement between two remote heterogeneous nodes. Such matter-matter entanglement is the fundamental networking resource as required for quantum state teleportation~\cite{Bennett1993PRL}, entanglement swapping~\cite{Briegel1998PRL}, and distributed quantum information processing~\cite{Gottesman1999Nat}. Establishing entanglement between a quantum processor and an ensemble-based quantum memory therefore remains an open challenge and constitutes an important step toward functional hybrid quantum networks.

Here, we demonstrate a prototype hybrid quantum network by generating entanglement between a trapped-ion (TI) processor node and a solid-state quantum memory (QM) node based on a rare-earth-ion ensemble. A spin-entangled photon emitted from the TI node is frequency-converted via a polarization-maintaining quantum frequency conversion (QFC) module and subsequently stored in, and retrieved on demand from a waveguide-integrated QM. We obtain a CHSH parameter~\cite{CHSH1969PRL} of $S = 2.328 \pm 0.054$, exceeding the classical bound by 6 standard deviations and thereby demonstrating entanglement between the two heterogeneous quantum systems separated by \SI{75}{\meter}.

\begin{figure*}[htbp]
    \centerline{\includegraphics[width=1\textwidth]{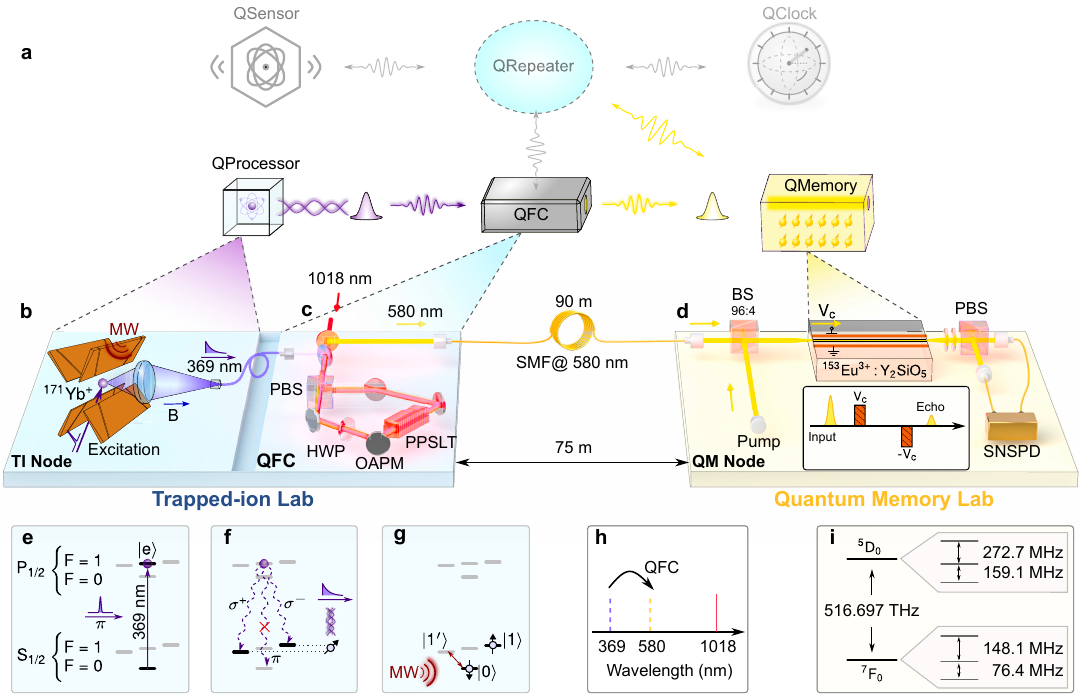}}
    \caption{\textbf{A prototype hybrid quantum network.}
     \textbf{a}, Schematic of the envisioned hybrid architecture, where heterogeneous quantum nodes are interconnected via flying photonic qubits. This work focuses on establishing entanglement between a quantum processor node based on trapped ion (TI) and a quantum memory (QM) node based on rare-earth ion ensemble, via a quantum frequency conversion (QFC) module. In this experiment, the two nodes are housed in separate laboratories spaced \SI{75}{\meter} apart and linked by a \hyphenSI{90}{\meter} single-mode fiber (SMF).
    \textbf{b}, The trapped-ion (TI) processor node. A single $\mathrm{^{171}Yb^{+}}$ ion is excited by a pulsed laser. Photons emitted along the quantization axis (defined by an external magnetic field) are collected by a high-numerical-aperture objective (NA = 0.64) and coupled into an SMF.
    \textbf{c}, The QFC module. Photons at \SI{369}{\nano\meter} are converted to \SI{580}{\nano\meter} via difference-frequency generation in a periodically poled stoichiometric lithium tantalate (PPSLT) waveguide, pumped at \SI{1018}{\nano\meter} (see panel \textbf{h}). A Sagnac interferometer configuration, comprising a PBS, OAPMs, and an intra-loop HWP, enables robust conversion independent of the input polarization.
    \textbf{d}, The QM node is based on a laser-written waveguide fabricated in an $\rm ^{153}Eu^{3+}\colon\!Y_2SiO_5$ crystal. An atomic-frequency comb (AFC) is prepared via optical pumping,  and on-chip electrodes enable electric-field control for on-demand readout of the $n^\text{th}$ AFC echo (Inset).
   \textbf{e--h}, Sequence detailing the ion-photon entanglement generation and the QFC process. 
   \textbf{i}, Relevant energy-level diagram of the QM. The quantum storage utilizes the polarization-independent ${}^{7}\mathrm{F}_0\leftrightarrow{}^{5}\mathrm{D}_0$ optical transition of site-2 $\mathrm{^{153}Eu^{3+}}$ ions in the $\mathrm{Y_2SiO_5}$ crystal. 
   Abbreviations: MW, microwave; SNSPD, superconducting nanowire single-photon detector; BS, beam splitter; HWP, half-wave plate; PBS, polarizing beam splitter; OAPM, off-axis parabolic mirror. 
    }
    \label{fig:network_arch}
\end{figure*}

\section*{A prototype hybrid quantum network architecture}
Fig.~\ref{fig:network_arch} presents the conceptual architecture of a hybrid quantum network, comprising heterogeneous nodes with distinct physical platforms and functionalities. Our experiment focus on three main components including the quantum process (QProcessor), the QFC module and the QM (Qmemory): the trapped-ion lab which includes TI node and the QFC module (Figs.~\ref{fig:network_arch}b--c),
and the QM node (Fig.~\ref{fig:network_arch}d). The two nodes, separated by \SI{75}{\meter}, are connected via a \hyphenSI{90}{\meter} single-mode fiber (SMF) serving as the quantum channel. Synchronization and classical communication, in contrast, are carried out through separate electrical signals.
A picosecond laser pulse at \SI{369}{\nano\meter} excites the single trapped $\mathrm{^{171}Yb^{+}}$ ion, which subsequently decays via spontaneous emission, returning to its ground state and emitting a photon whose polarization is entangled with the ion spin~\cite{Maunz2007natphy,Moehring2007,Matsukevich2008PRL,Vittorini2014PRA,Hucul2015natphy}.
This \hyphenSI{369}{\nano\meter} photon is collected by an objective and then routed to the QFC module, where its wavelength is converted to \SI{580}{\nano\meter} while preserving its polarization-encoded quantum state. The converted photon is transmitted through the \hyphenSI{90}{\meter} SMF to the QM node, where it is stored and later retrieved on demand.

The TI node employs a high-optical-access ion trap~\cite{he2021RSI}, 
enabling efficient photon collection via an objective with a NA of 0.64, 
achieving a collection efficiency of approximately 10\% and subsequent coupling into a \hyphenSI{3}{\meter} SMF.
The QFC module implements a difference-frequency generation (DFG) process: as illustrated in Fig.~\ref{fig:network_arch}h, 
a \hyphenSI{1018}{\nano\meter} pump laser drives low-noise, polarization-maintaining conversion from \SI{369}{\nano\meter} to \SI{580}{\nano\meter} to match trapped ion with the solid-state QM based on $\rm ^{153}Eu^{3+}$ ensemble.
The QM node is based on a laser-written depressed-cladding optical waveguide fabricated on a 0.2\% doped $\rm ^{153}Eu^{3+}\colon\!Y_2SiO_5$ crystal, which supports waveguiding for arbitrary polarization states. To enable polarization-independent storage, we utilize site-2 $\rm Eu^{3+}$ ions, whose ${}^{7}\mathrm{F}_0\leftrightarrow{}^{5}\mathrm{D}_0$ transition exhibits balanced absorption for light polarized along the crystal's $D1$ and $b$ axes~\cite{konz2003temperature}. Relative to our previous work employing 0.1\% doped $\rm ^{151}Eu^{3+}\colon\!Y_2SiO_5$~\cite{zhu2022demand}, 
the increased doping concentration and the switch to the $\rm ^{153}Eu^{3+}$ isotope yield a $\sim$2-fold enhancement in storage efficiency and a $\sim$5-fold increase in operational bandwidth. However, higher doping concentrations also intensify flip-flop processes and instantaneous spectral diffusion, which could degrade the contrast of the AFC structure, the effective AFC lifetime, and the effective AFC coherence time~\cite{liu2024NC, Zhu2026natpho}. In the present demonstration, these effects are negligible, as the AFC storage time is relatively short (\si{\micro\second}). 
On-demand readout is achieved via on-chip electrodes that apply control electric pulses, following the Stark-modulated atomic frequency comb (SMAFC) protocol \cite{horvath2021noise, liu2020demand}. The device is cooled to \SI{3}{\kelvin} in a closed-cycle cryostat (Montana Instruments). 

\section*{Ion-photon entanglement}
The entanglement between the ion's spin and the polarization of the emitted \hyphenSI{369}{\nano\meter} photon is generated via the protocol sketched in Fig.~\ref{fig:network_arch}e-f~\cite{Maunz2007natphy,Moehring2007}.
Initially, the ion is optically pumped to the state
$\lvert 0 \rangle \equiv \mathrm{S}_{1/2}\lvert F=0, m_F=0\rangle$.
Subsequently, a single picosecond laser pulse with $\pi$ polarization is applied to excite the ion to the state
$\lvert e \rangle \equiv \mathrm{P}_{1/2}\lvert F=0, m_F=0\rangle$.
Owing to selection rules, spontaneous decay from the excited state $\lvert e\rangle$ populates the $\mathrm{S_{1/2}}$ manifold with equal probability (1/3 each) after excitation:
$\left|0 \right>$,
$\left|1'\right> \equiv \left|F=1, m_F=-1\right>$, and
$\left|1\right>  \equiv \left|F=1, m_F=1\right>$.
The hyperfine states $\left|1\right\rangle$, $\left|1'\right\rangle$, and $\left|0\right\rangle$ are correlated with the photon polarizations $\sigma^-$, $\sigma^+$, and $\pi$, respectively.
These photons (the flying qubits) are collected by an objective aligned parallel to the quantization axis, which is defined by a magnetic field of approximately \SI{4}{\gauss}, and are then coupled into a SMF. In this geometry, the $\sigma^+$ and $\sigma^-$ components are coupled efficiently into the fiber mode, whereas the $\pi$-polarized photons are suppressed by the collection geometry and are further rejected by the spatial filtering inherent to the SMF mode profile~\cite{PRL.124.110501}.
Taking $t=0$ as the time when the picosecond pulse prepares the atom in $\ket{e}$, consider the conditional state generated by a photon-emission event at time $t_0$. The two relevant decay paths are $\ket{e}\rightarrow\ket{1'}+\sigma^+$ and $\ket{e}\rightarrow\ket{1}+\sigma^-$. The two emitted photon modes differ not only in polarization but also in center frequency by the ground-state Zeeman splitting $\omega$. Keeping this frequency degree of freedom implicit in the photon-mode labels, the resulting ion-photon state conditioned on this emission event can be written compactly as
\begin{equation}
    |\Psi\rangle=\frac{1}{\sqrt{2}}\left(|1'\rangle|+\rangle+e^{-i\phi}|1\rangle|-\rangle\right),
    \label{eq1}
\end{equation}
where $\ket{+}$ ($\ket{-}$) denotes the photon mode associated with the $\sigma^+$ ($\sigma^-$) transition. 
The relative phase is determined by the emission time measured from the excitation pulse, $\phi=\omega t_0+\phi_0$, where $\omega=2\pi\times \SI{11.22}{\mega\hertz}$ and $\phi_0$ is a fixed offset set by the chosen photon-mode phase convention and by the phase reference of the MW pulse used for ion-state analysis~\cite{Vittorini2014PRA}.
A photon click is recorded only later, at $t_c=t_0+T_{\mathrm{delay}}$, where $T_{\mathrm{delay}}$ is the fixed delay for both direct detection and detections after the QFC or the QM. The click therefore heralds the successful event and provides the timing information needed to track $\phi$ for each trial (see Section~10 of the Supplementary Information).

\begin{figure*}[htbp]
    \centerline{\includegraphics[width=1\textwidth]{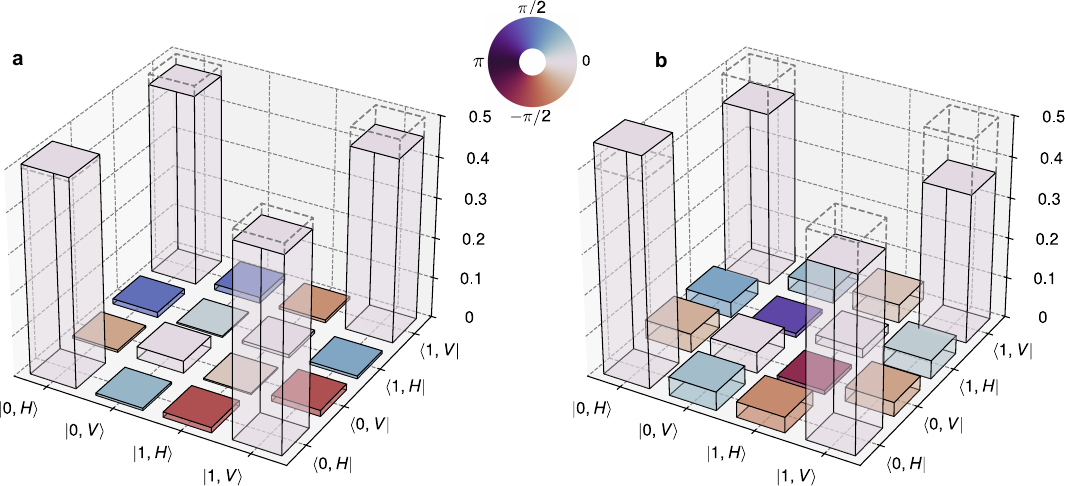}}
    \caption{
    \textbf{Reconstructed density matrices of the ion-photon entangled state.} Matrices are shown before (\textbf{a}) and after the QFC (\textbf{b}), as obtained via quantum state tomography. The corresponding fidelities are $(95.46 \pm 0.24)\%$ (\textbf{a}) and $(89.1 \pm 1.9)\%$ (\textbf{b}) to the target Bell state. The dashed lines denote the expected results for ideal Bell states.
    }
    \label{fig:ion_photon_matrix}
\end{figure*}

Upon such a herald, the ion state is analyzed with a pre-calibrated MW waveform generated by an arbitrary waveform generator (AWG). This waveform maps the population from $\ket{1'}$ onto $\ket{0}$ and has a fixed phase chosen for the desired ion-analysis basis; because it is triggered by the heralding click, the emission-time-dependent phase in Eq.~\ref{eq1} appears as a fixed offset in the triggered MW reference frame and is included in the waveform calibration (Fig.~1g). The ion-photon entanglement is then characterized by quantum state tomography (QST; see Section~9 of the Supplementary Information), with the $\sigma^+$ and $\sigma^-$ photonic components mapped onto the $\ket{H}$ and $\ket{V}$ polarization states, respectively, using a set of wave plates before the QST. 
The reconstructed density matrix is shown in Fig.~\ref{fig:ion_photon_matrix}. Based on 62,723 measurement trials, we obtain a fidelity of $(95.46 \pm 0.24)\%$ with respect to the target Bell state, where the uncertainty is estimated using a Monte Carlo simulation~\cite{Huang2025Sciadv,DeGreve2013NC,Gao2012Nat} (Section~11 in Supplementary Information). 

The QFC module subsequently converts the wavelength of the flying qubits from \SI{369}{\nano\meter} to \SI{580}{\nano\meter}, via DFG driven by a \hyphenSI{4}{\watt}, \hyphenSI{1018}{\nano\meter} pump laser. The QFC scheme is designed to preserve photonic polarization-a critical requirement for maintaining entanglement-while simultaneously suppressing spontaneous parametric down-conversion noise and Raman scattering. The conversion is implemented using a crossband Sagnac interferometer architecture~\cite{Ikuta_2018,Arenskotter2023,PRL.124.010510,Yao26col} (see Methods). After propagation through the \hyphenSI{90}{\meter} SMF to the QM lab, detection of the output flying qubits yield an end-to-end QFC efficiency of 0.076\% (see in Methods). Based on 2,714 measurement trials, the fidelity of the ion-photon entangled state after frequency conversion is $(89.1 \pm 1.9)\%$.

\begin{figure*}[ht]
    \centerline{\includegraphics[width=1\textwidth]{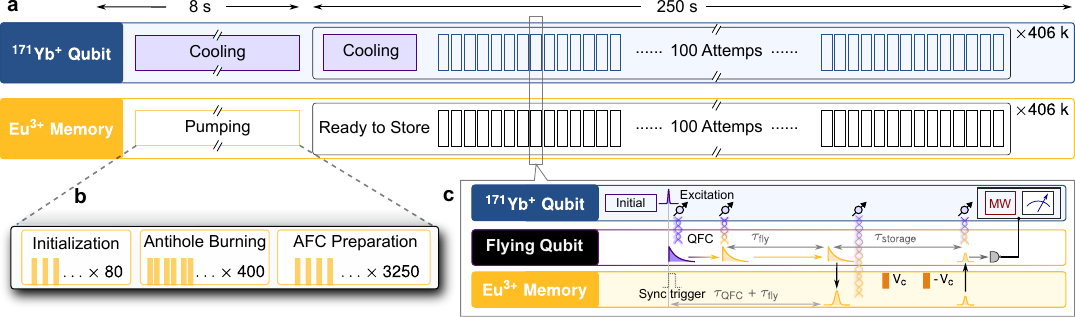}}
    \caption{
    \textbf{Timed sequence for entangling the TI node and the QM node.}
    \textbf{a}, Overall time sequence. Each experimental cycle comprises an \hyphenSI{8}{\second} QM preparation phase followed by a \hyphenSI{250}{\second} entanglement storage phase. During QM preparation, the trapped ion is maintained under continuous laser cooling. The storage phase is divided into multiple frames, each containing \SI{100}{\micro\second} of ion-laser cooling followed by 100 sequential storage attempts.
    \textbf{b}, QM preparation. This process involves three stages: initialization, antihole burning, and atomic-frequency-comb (AFC) preparation.
    \textbf{c}, Single entanglement-generation attempt. The sequence begins with ion-qubit initialization (\SI{2.5}{\micro\second}, Fig.~\ref{fig:network_arch}\textbf{e}). A \hyphenSI{369}{\nano\meter} laser pulse then excites the ion to generate ion-photon entanglement, producing a flying qubit at \SI{369}{\nano\meter} (Fig.~\ref{fig:network_arch}\textbf{f}). The emitted photon is converted to \SI{580}{\nano\meter} by the QFC module and transmitted through the optical fiber connecting the two labs, introducing a delay of
    $\tau_\mathrm{fly}=\SI{484}{\nano\second}$.
    Finally, the \hyphenSI{580}{\nano\meter} photon is stored and on-demand retrieved from the QM node. The photon-detection signal in the QM node is sent back to the TI node to herald whether a ion-state readout should be performed.
    }
    \label{fig:time_seq}
\end{figure*}

\section*{Heterogeneous entanglement between TI node and QM node}
The complete timing sequence for generating entanglement between the TI node and the QM node is depicted in Fig.~\ref{fig:time_seq}a. Each experimental cycle begins with an \hyphenSI{8}{\second} preparation phase for the QM (Fig.~\ref{fig:time_seq}b), during which the TI is kept under continuous Doppler cooling. This is followed by a \hyphenSI{250}{\second} entanglement-storage phase, consisting of 0.406 millions repeated storage frames. Every frame starts with \SI{100}{\micro\second} of ion cooling and then executes 100 entanglement-generation attempts.
The workflow of a single attempt is illustrated in Fig.~\ref{fig:time_seq}c: Firstly, the ion qubit is initialized to the $\left|0\right\rangle$ state via optical pumping, which takes \SI{2.5}{\micro\second} (Fig.~\ref{fig:network_arch}e). A single \hyphenSI{369}{\nano\meter} picosecond pulse then excites the ion, generating entanglement between its Zeeman sublevels and the polarization of the emitted \hyphenSI{369}{\nano\meter} flying qubit (Fig.~\ref{fig:network_arch}f).
The flying qubit is then directed to the QFC module to establish the entanglement between the ion qubit and the \hyphenSI{580}{\nano\meter} photon.
After conversion, the \hyphenSI{580}{\nano\meter} photon propagates through the \hyphenSI{90}{\meter} SMF, reaching the QM lab after a flight time  $\mathrm{\tau_{fly}}=\SI{484}{\nano\second}$. Finally, it is stored in the QM node via SMAFC protocol ~\cite{liu2020demand, horvath2021noise} with storage times actively controlled by electric pulses. The AFC structure is prepared with a periodicity of \SI{2}{\mega\hertz}, which enables on-demand readout at discrete times given by $n \times \tau_{\text{echo}}$, where $\tau_{\text{echo}} = \SI{0.5}{\micro\second}$. The integer $n$ is determined by the timing of the second electric pulse. The details of the QM preparation are presented in Section~5 in the Supplementary information.

We begin by assessing the QM performance using weak coherent pulses with a mean photon number of 0.12 (Supplementary Fig.~6). The storage time can be switched from \SI{0.5}{\micro\second} and \SI{1}{\micro\second} by applying the second control electric pulse within the time window from \SI{0.5}{\micro\second} to \SI{1}{\micro\second}. For input states $\left|H\right\rangle$ and $\left|V\right\rangle$, the \hyphenSI{0.5}{\micro\second} (\hyphenSI{1}{\micro\second}) SMAFC protocol achieves internal efficiencies of $(43.3\pm0.2)\%$ ($(31.0\pm0.1)\%$) and $(39.8\pm0.2)\%$ ($(28.9\pm0.1)\%$), respectively.  After accounting for optical losses, the total device efficiencies  at \SI{1}{\micro\second} are $(19.5\pm0.1)\%$ for $\left|H\right\rangle$ and $(18.3\pm0.1)\%$ for $\left|V\right\rangle$). We then assess the memory's quantum performance by storing encoded polarization qubits for \SI{1}{\micro\second}. The storage fidelities for the states $\left|H\right\rangle$, $\left|V\right\rangle$, $\left|H\right\rangle+\left|V\right\rangle$, and $\left|H\right\rangle+i\left|V\right\rangle$ reach $(99.97\pm0.01)\%$, $(99.90\pm0.01)\%$, $(99.81\pm0.02)\%$, and $(99.34\pm0.03)\%$, respectively. The high fidelity directly result from the inherently noise-free nature of the SMAFC protocol.

\begin{figure*}[htbp]
    \centerline{\includegraphics[width=1\textwidth]{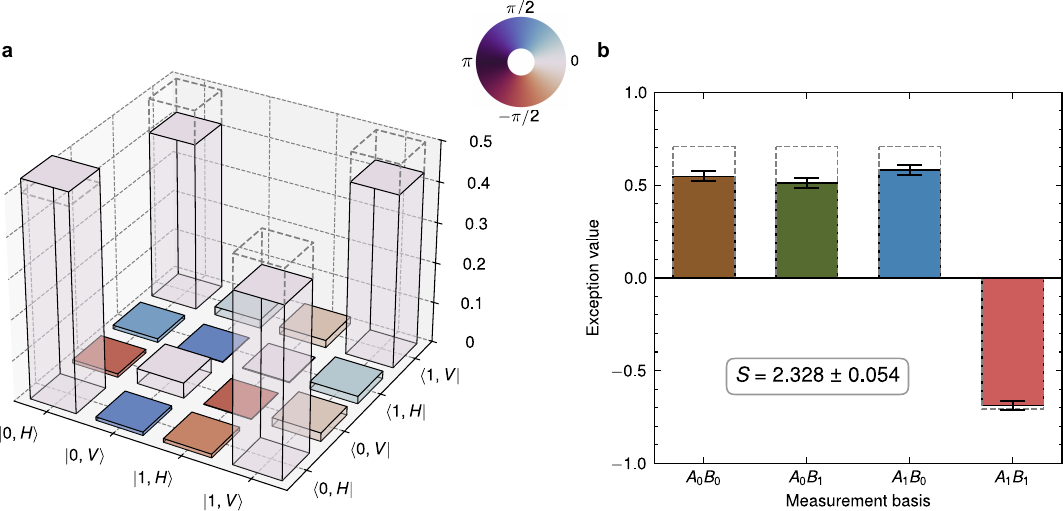}}
    \caption{\textbf{Characterization of the TI-QM entangled state after on-demand storage for \SI{1}{\micro\second} using the SMAFC protocol.} \textbf{a}, Density matrix reconstructed via quantum state tomography, yielding a fidelity of (89.2$\pm$2.3)\%. \textbf{b}, test of the CHSH form of the Bell inequality. Measurements in four basis settings give a CHSH parameter~\cite{CHSH1969PRL} of $S = 2.328 \pm 0.054$, based on 3,634 trials.
    }
    \label{fig:TI_QM_matrix}
\end{figure*}

To ensure spectral matching between the frequency-converted photons and the QM, the \hyphenSI{1018}{\nano\meter} pump laser for QFC was frequency-stabilized using the Pound–Drever–Hall (PDH) technique, achieving a stabilized linewidth of approximately \SI{300}{\kilo\hertz}. 
Spectral matching is optimized by aligning the center frequency of the reference converted beam at \SI{580}{\nano\meter} and the absorption band of the QM (Methods). After this, we experimentally obtained a bandwidth-matching efficiency of approximately $(74\pm11)\%$ using \hyphenSI{580}{\nano\meter} flying qubits, close to the theoretical limit of $74.34\%$ for $\Delta f=0$ (Section~7 in Supplementary Information). 
The internal storage efficiency for spin-entangled photons matches that measured with weak coherent pulses, reaching average efficiencies of $41.6\%$ at \SI{0.5}{\micro\second} and $30.0\%$ at \SI{1}{\micro\second}. 
These results establish a record efficiency for entanglement storage in a solid-state system, which was previously limited to a record value of 27\%~\cite{Liu_2021,Lago-Rivera2021Nat,Zhu2026natpho}, achieved here within a waveguide-integrated architecture. 
The overall end-to-end efficiency-from the \hyphenSI{369}{\nano\meter} input of QFC module to final retrieval at \SI{580}{\nano\meter} is approximately 0.011\% at a storage time of \SI{1}{\micro\second} (Methods). This corresponds to a TI-QM entanglement generation rate of \SI{0.2}{\hertz}. To suppress noise, a \hyphenSI{30}{\nano\second} photon-detection window is applied in the temporal domain (Section~4 in Supplementary information). Combined with the QM, which acts as a \hyphenSI{48.2}{\mega\hertz} narrow-band spectral filter, the total noise count rate $N_{\mathrm{noise}}$ is reduced to \SI{0.007}{\hertz}, which predominantly originates from the dark noise of SNSPD and the environment (Section~4 in Supplementary Information). $N_{\mathrm{noise}}$ is measured by blocking the photon emission from the trapped ion while keeping all other experimental conditions identical to the signal acquisition mode. Consequently, the signal-to-noise ratio (SNR) for the TI-QM system reaches 28:1, defined as the ratio of the retrieved single-photon count rate ($N_{\mathrm{signal}}$) to the noise count measured with the photon emission from the trapped ion turned off ($N_{\mathrm{noise}}$). The generated TI-QM entanglement was verified by QST of the joint ion-spin and retrieved-photon states. Based on 1,780 successful trials, the reconstructed density matrix (Fig.~\ref{fig:TI_QM_matrix}a) yields a fidelity of (89.2 $\pm$ 2.3)\%. 
The dominant contributions to the entanglement infidelity are from the QFC module(3.1\%), the pulse excitation process (3.3\%), detection error caused by dark noise (2.7\%) and state preparation and measurement error (0.7\%). Other infidelity sources contribute about 0.8\% (Section~8 in Supplementary Information). This fidelity is consistent with that of the ion-photon state measured prior to storage, confirming the stable operation of the hybrid quantum network and the reliable storage of the QM.
Using the reconstructed matrices shown in Fig.~\ref{fig:time_seq}b and Fig.~\ref{fig:TI_QM_matrix}a, we obtain a relative fidelity of ($97.3\pm2.2$)\% between the input and output ion‑photon entangled states for the QM. This value provides a conservative upper bound of 2.7\% on any infidelity introduced by the storage process for photons from the TI, with dark noise being the dominant contribution to the overall infidelity.

\section*{Violation on Bell inequality between heterogeneous quantum nodes}
The violation of Bell inequality provides the definitive benchmark for certifying entanglement and underpins numerous quantum-information protocols~\cite{Pironio2010Nat,Nadlinger2022Nat,PhysRevLett.132.150604}. We therefore perform a test of the CHSH form of the Bell inequality, formally quantified by the CHSH parameter~\cite{CHSH1969PRL}
\begin{equation}
    S=\langle A_0B_0 \rangle+\langle A_0B_1 \rangle+\langle A_1B_0 \rangle-\langle A_1B_1 \rangle,
    \label{eq2}
\end{equation}
where $\langle A_{0,1}B_{0,1} \rangle$ denotes the expectation value of the product of measurement outcomes at the TI node and the QM node. The spin qubit in the TI node is measured in ${|0\rangle\pm|1\rangle}/{\sqrt{2}}$ basis ($A_0$) and $|0\rangle/|1\rangle$ basis ($A_1$), while the photonic polarization qubit retrieved from the QM is measured in $|H\rangle/|V\rangle$ basis ($B_0$) and $(|H\rangle\pm|V\rangle)/\sqrt{2}$ basis ($B_1$). With outcomes assigned values $\pm 1$, any local hidden-variable theory is bounded by $|S| \leq 2$. For our TI-QM entangled state, we obtain $S = 2.328 \pm 0.054$ (Fig.~\ref{fig:TI_QM_matrix}b), violating the classical limit by more than 6 standard deviations. This result unambiguously confirms the presence of high-quality entanglement between the two distinct quantum nodes.

\section*{Conclusion and discussion}\label{sec6}

In summary, we have demonstrated a prototype hybrid quantum network that realizes entanglement between a trapped-ion node and a solid-state quantum memory node operating at distinct wavelengths, enabling a $6\sigma$ violation of the Bell inequality. This successful integration of a heterogeneous single-atom processing node with an atomic-ensemble memory node establishes a key hardware foundation for scalable quantum networks.

The performance of this architecture can be further enhanced along several pathways. For the TI node, the photon collection efficiency could be increased from the current 10\% to exceeding 70\% by combining cavity with trapped ion~\cite{Takahashi2020PRL, 2021SchuppPRXQ,fang2025PRA} and the number of qubits in this architecture can be extended by shuttling different ions into the cavity~\cite{Canteri2025PRL}. 
The QFC internal efficiency, currently 0.7\%, is mainly limited by the commercially available  PPSLT crystal used in this work, whose \SI{7.36}{\micro\meter} poling period enables third-order, rather than first-order, QPM for the 369-nm-to-580-nm conversion. Shorter‑period first‑order poled waveguides could achieve 30\% efficiency for UV-to-telecom QFC~\cite{yang2026highQFC} which would enable long‑distance deployment of hybrid quantum networks. 
For the QM node, storage efficiency exceeding 80\% is achievable by employing an impedance-matched optical cavity~\cite{Meng2026natpho}, while operating at a specific magnetic field could extend the storage time into the minute-scale regime~\cite{lv2025Minute-Scale}. 
Incorporating these improvements still represents significant engineering challenges, but achieving this could transform this hybrid architecture into a practical platform for enabling novel quantum computing architectures~\cite{PRL.127.140503} and distributed quantum networks~\cite{tissot2025,Gu2025npj,Cussenot2025prl}.
\renewcommand{\thefigure}{\arabic{figure}}
\captionsetup[figure]{name=Extended Fig.}
\setcounter{figure}{0}

\renewcommand{\thetable}{\arabic{table}}
\captionsetup[table]{name=Extended Table}
\setcounter{table}{0}

\section*{Methods}\label{method}

\subsection*{Fabrication of the QM device}\label{QMdevice}

\begin{figure*}[h]
    \centerline{\includegraphics[width=\textwidth]{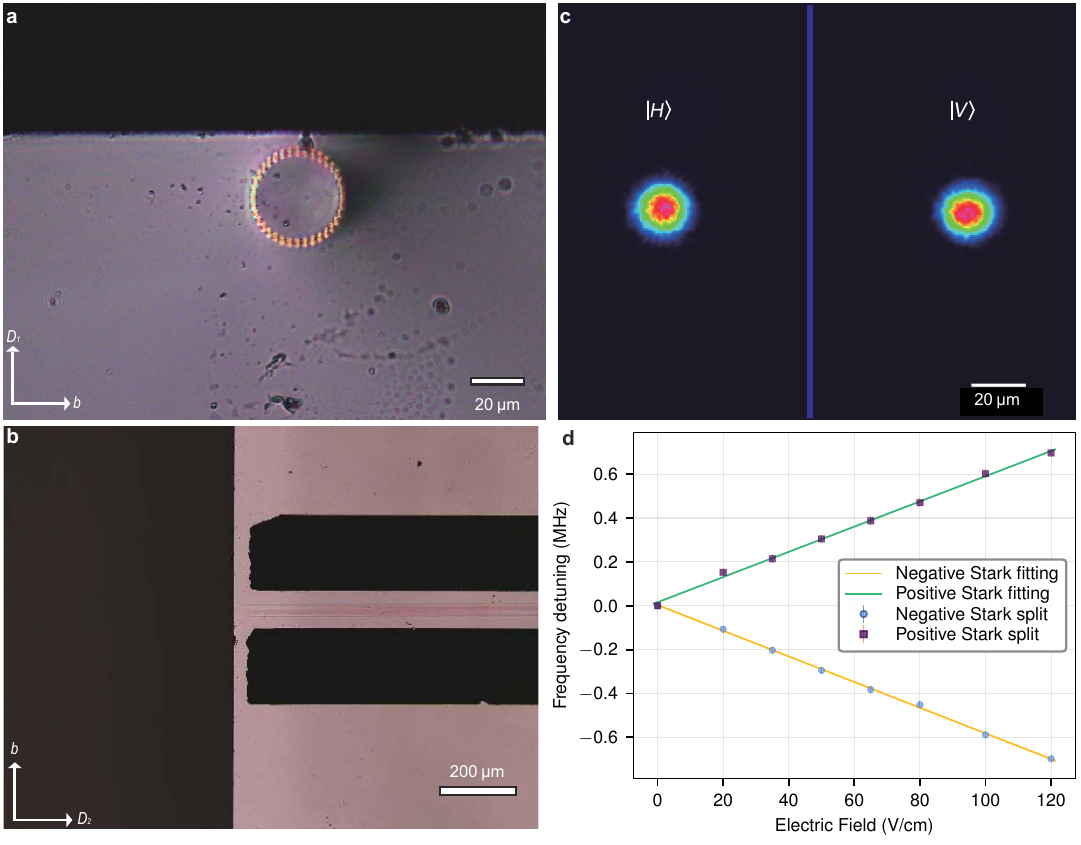}}
    \caption{\textbf{Fabrication and characterization of the integrated quantum memory device.}
    \textbf{a}, Front and \textbf{b}, top view of the laser-written depressed-cladding waveguide. The waveguide (visible between the two black electrodes) is aligned along the crystal's $D_2$ axis. Dark spots are due to surface contamination. Scale bars: \SI{20}{\micro\meter} (\textbf{a}) and \SI{200}{\micro\meter} (\textbf{b}). \textbf{c}, Beam profile at the waveguide output for the $|H\rangle$ and $|V\rangle$ polarization states; scale bar: \SI{20}{\micro\meter}. 
    \textbf{d}, Stark splitting of the doped $\mathrm{^{153}Eu^{3+}}$ ions  as a function of the applied electric field. The average Stark splitting is \SI{5.80 \pm 0.09}{\kilo\hertz\per\volt\per\centi\meter}.
}
    \label{fig:QMwaveguide}
\end{figure*}

The quantum memory is fabricated on a $\rm ^{153}Eu^{3+}:Y_2SiO_5$ crystal (0.2\% doping concentration) with dimensions of $4\times25.6\times5$ \si{\cubic\milli\meter} ($D_1\times D_2\times b$). We employ a commercial femtosecond-laser micromachining (FLM) system (WOPhotonics) to inscribe a depressed-cladding optical waveguide with a diameter of \SI{30}{\micro\meter}, consisting of 36 concentric tracks (Extended Fig.~\ref{fig:QMwaveguide}a). The waveguide core is positioned \SI{27}{\micro\meter} beneath the crystal surface to ensure good coupling with the on-chip electrodes.
The fabrication uses a \hyphenSI{1030}{\nano\meter} laser source with the following parameters: pulse width \SI{210}{\femto\second}, repetition rate \SI{201.9}{\kilo\hertz}, and pulse energy \SI{70}{\nano\joule}. The laser beam, polarized along the $D_2$ axis, is focused through a $100\times$ objective (NA = 0.7) and illuminates the crystal along the $D_1$ axis. The laser beam moves along the $D_2$ axis at a speed of \SI{5}{\milli\meter\per\second}. The fabricated waveguide exhibits a nearly Gaussian output mode (Extended Fig.~\ref{fig:QMwaveguide}c), with fitted FWHM dimensions of \SI{11.4}{\micro\meter} $\times$ \SI{12.5}{\micro\meter} for $|H \rangle$ and \SI{12.0}{\micro\meter} $\times$ \SI{12.4}{\micro\meter} for $|V \rangle$ polarization states.
The measured insertion losses are \SI{0.50}{\deci\bel\per\centi\meter} ($|H\rangle$) and \SI{0.47}{\deci\bel\per\centi\meter} ($|V\rangle$). The total optical path efficiency $T_\mathrm{QM}$ from the cryostat front to the collection into a SMF is 63.0\% for $|H\rangle$ and 63.4\% for $|V\rangle$.

On-chip electrodes (\hyphenSI{50}{\nano\meter} Ti / \hyphenSI{150}{\nano\meter} Au) with a spacing of \SI{100}{\micro\meter} are patterned on the crystal surface via UV photolithography and electron-beam evaporation (Extended Fig.~\ref{fig:QMwaveguide}b).
When an electric field is applied along the $b$ axis, the $\mathrm{^{153}Eu^{3+}}$ ions experience a linear Stark shift, splitting into two sub-ensembles with opposite frequency shifts. Spectral-hole-burning measurements on site-2 ions give shift rates of \SI[retain-explicit-plus]{+5.74 \pm 0.07}{\kilo\hertz\per\volt\per\centi\meter} and \SI{-5.85 \pm 0.12}{\kilo\hertz\per\volt\per\centi\meter} (Extended Fig.~\ref{fig:QMwaveguide}d), consistent with values reported for $\rm ^{151}Eu^{3+}$ ions in the same site \cite{zhu2022demand}.

\subsection*{Detail of QFC optical path and signal rate}\label{detailofQFC}
\begin{figure}[h]
    \centerline{\includegraphics[width=0.47\textwidth]{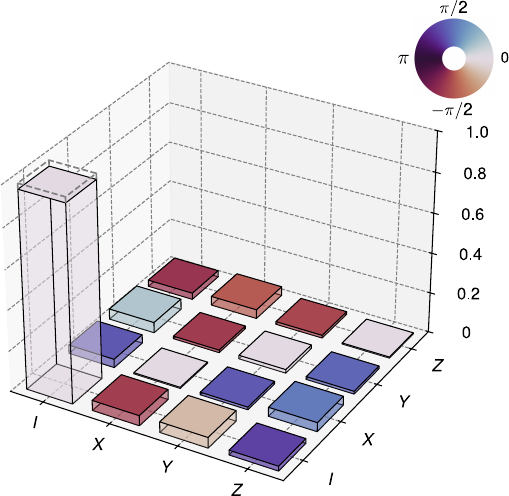}}
    \caption{
    \textbf{The reconstructed process matrix of the QFC module and the \hyphenSI{90}{\meter} SMF.}
    The process matrix is reconstructed via quantum process tomography using a tomographically complete set of six input polarization states: $|H\rangle, |V\rangle, |D\rangle, |A\rangle, |R\rangle$, and $|L\rangle$. The indices $I, X, Y, Z$ denote the identity and Pauli operators, corresponding to the vacuum, $\ket{H}/\ket{V}$, $\ket{D}/\ket{A}$, and $\ket{R}/\ket{L}$ bases, respectively.
    The process fidelity relative to the ideal identity matrix is $(97.32\pm0.14)$\%, obtained with a measurement time of \SI{3}{\minute}.}
    \label{fig:process_matrix}
\end{figure}

The QFC module employs a Sagnac loop configuration to realize the DFG process to convert photons from \SI{369}{\nano\meter} to \SI{580}{\nano\meter} by using a \hyphenSI{1018}{\nano\meter} pump laser and type-0 quasi-phase matching (QPM)~\cite{Yao26col}. As shown in Fig.~\ref{fig:network_arch}c, the Sagnac loop is placed downstream of the polarizing beam splitter (PBS), and then be divided into clockwise (CW) and counter-clockwise (CCW) sub-loops according to the propagation directions of the incident light after passing through the PBS. Specifically, the CW sub-loop is responsible for converting the $|V\rangle$ state component of the incident light, while the CCW sub-loop undertakes the conversion of the $|H\rangle$ state polarization component. This configuration enables polarization-independent, high-fidelity, and high-stability DFG. The PPSLT crystal has dimensions of \SI{10}{\milli\meter} $\times$ \SI{5}{\milli\meter} $\times$ \SI{0.5}{\milli\meter}, with a \SI{7.36}{\micro\meter} poling period (OXIDE Corporation). A waveguide with \SI{10}{\milli\meter} length is also fabricated with the FLM system. The crystal temperature was stabilized at \SI{109.7}{\celsius}, thereby realizing third-order QPM.
To achieve simultaneous coupling of these three wavelengths, off-axis parabolic mirrors (OPAM) are employed (Fig.~\ref{fig:network_arch}c).
Through optimization of the waveguide structure and optical alignment,
we achieve coupling efficiencies of $\eta_{\mathrm{369}}= 0.708$ for \SI{369}{\nano\meter} and $\eta_{\mathrm{1018}}=0.6$ for \SI{1018}{\nano\meter}. Meanwhile, the \hyphenSI{580}{\nano\meter} photons generated by the QFC module are separated using a dichroic mirror and coupled into the \hyphenSI{90}{\meter} SMF.
Small polarization-dependent QFC efficiency difference is balanced by fine-tuning the pump powers in the two sub-loops of Sagnac loop. As shown in Extended Fig.~\ref{fig:process_matrix}, we achieve an overall process matrix fidelity of ($97.32\pm0.14$)\%, including QFC module and \hyphenSI{90}{\meter} SMF.

Throughout the experiment, three distinct stages of entanglement generation rates are defined: $R_\mathrm{369}$, $R_\mathrm{580}$ and $R_\mathrm{TI-QM}$, where $R_\mathrm{369}$ denotes the entanglement generation rate between TI node and \hyphenSI{369}{\nano\meter} flying qubits, $R_\mathrm{580}$ represents the entanglement generation rate between TI node and \hyphenSI{580}{\nano\meter} flying qubit, and $R_\mathrm{TI-QM}$ is corresponds to the entanglement generation rate between TI node and QM node.

$R_\mathrm{369}$ can be calculated by $R_\mathrm{369}=2/3P_\mathrm{\pi}R_\mathrm{exp}P_\mathrm{S_{1/2}}Q_\mathrm{E369}T_\mathrm{fib1}T_\mathrm{opt}E_\mathrm{obj}$, where $P_\mathrm{\pi}=0.960$ is the excitation probability. $R_\mathrm{exp1}=\SI{250}{\kilo\hertz}$ is the repetition rate of entanglement generation attempts, where each attempt consists of a \SI{1}{\micro\second} cooling, a \SI{2.5}{\micro\second} pumping, and a \SI{0.5}{\micro\second} delay time. This delay time corresponds to the flying time required for flying qubits to reach the detectors after a single-pulse excitation command is issued. Only when photons arrive at the detectors and a valid photon-detection event is confirmed will the ion-state readout be performed.
Additional parameters are defined as follows:
$P_\mathrm{S_{1/2}}=0.995$ is the branching ratio of $P_{1/2}-S_{1/2}$,
$Q_\mathrm{E369}=0.35$ is the quantum efficiency of PMTs,
$T_\mathrm{fib1}=0.27$ is the fiber coupling efficiency of \hyphenSI{3}{\meter} SMF,
$T_\mathrm{opt}=0.9$ is the photon transmission through imaging optical path and $E_\mathrm{obj}$ is the collection efficiency of 0.64-NA objective.
Under these parameters, the generation rate of entanglement between TI node and \hyphenSI{369}{\nano\meter} flying qubits is $R_\mathrm{369}=\SI{1352}{\hertz}$.

For $R_\mathrm{580}$, the increase in the delay time leads to a decrease in the the repetition rate of entanglement generation attempts. The delay time is \SI{1.66}{\micro\second}, which results $R_\mathrm{exp2}=\SI{194}{\kilo\hertz}$ for entanglement between TI node and \hyphenSI{580}{\nano\meter} flying qubits. The end-to-end efficiency of the QFC module defined as $\eta_\mathrm{QFC}=\eta_{\mathrm{369}}\eta_{\mathrm{conv}}T_{\mathrm{580}}T_{\mathrm{fib2}}\eta_\mathrm{AOM}$ represents a key performance metric that constrains the value of $R_\mathrm{580}$,
where $\eta_{\mathrm{conv}}=0.007$ is the internal conversion efficiency under a pump laser power of \SI{2}{\watt} per sub-loop,
$T_{\mathrm{580}}=0.478$ is the optical path transmittance at \SI{580}{\nano\meter} of QFC module,
$T_{\mathrm{fib2}}=0.4$ is the coupling efficiency and transmittance of \hyphenSI{90}{\meter} SMF and $\eta_\mathrm{AOM}=0.8$ is the diffraction efficiency of AOM which is used for noise suppression (see more detail about this AOM in Section~4 in Supplementary information).
To get higher entanglement generation rate, SNSPDs with quantum efficiency $Q_\mathrm{E580}=0.8$ are used for detection of \hyphenSI{580}{\nano\meter} photons.
Collectively, these efficiencies yield
$R_\mathrm{580}=\frac{Q_\mathrm{E580}}{Q_\mathrm{E369}}\times R_\mathrm{369}\times\frac{R_\mathrm{exp2}}{R_\mathrm{exp1}}\times\eta_{\mathrm{QFC}}=\SI{1.8}{\hertz}$.

For $R_\mathrm{TI-QM}$, the delay is further increased to \SI{2.66}{\micro\second} and the repetition rate of entanglement generation attempts decrease to $R_\mathrm{exp3}=\SI{162}{\kilo\hertz}$. The end-to-end efficiency of QM node is defined as $\eta_\mathrm{QM}=\eta_\mathrm{bw}\eta_\mathrm{storage}$, where
$\eta_\mathrm{storage}$ is the QM's device efficiency of \hyphenSI{1}{\micro\second} storage time for $|H\rangle$ ($|V\rangle$) considering optical path loss, which is $(19.5\pm0.1)\%$ ($(18.3\pm0.1)\%$) as shown in Supplementary information section~5,
$\eta_\mathrm{bw}=0.74$ is the bandwidth matching efficiency between \hyphenSI{580}{\nano\meter} flying qubits and QM (see section~7 Supplementary information). To align the center frequencies of the QFC output and the QM, a \hyphenSI{1}{\milli\watt} reference beam resonant with the ion $\left|0\right\rangle \leftrightarrow \left|e\right\rangle$ transition was sent through the QFC module, producing approximately \SI{10}{\micro\watt} of converted \hyphenSI{580}{\nano\meter} light. We tuned the frequency of the \hyphenSI{1018}{\nano\meter} pump laser while monitoring the converted reference beam. The frequency offset of the converted light was characterized by heterodyne beating with a hole-burning laser referenced to the center of the \hyphenSI{48.2}{\mega\hertz} QM absorption band. The resulting beat note enabled alignment of the center frequency to within \hyphenSI{1}{\mega\hertz} accuracy. 
These efficiencies yield $R_\mathrm{TI-QM}=R_\mathrm{580}\times\frac{R_\mathrm{exp3}}{R_\mathrm{exp2}}\times\eta_\mathrm{QM}=\SI{0.2}{\hertz}$.
The overall efficiency of QFC module and QM node $\eta=\eta_\mathrm{QFC}\eta_\mathrm{QM}$ is about 0.011\%.
All efficiency components contributing to $\eta$ are listed in Extended Table~\ref{tab2}.

\begin{table}[htbp]
    \caption{Photon efficiency for each stage of QFC module and QM node.
    For the internal conversion efficiency $\eta_{\mathrm{conv}}$, the pump power (given in parentheses) is adjusted separately to balance conversion efficiencies of $\left|H\right\rangle$ and $\left|V\right\rangle$.
    }
    \begin{tabular*}{1\textwidth}{@{\extracolsep{\fill}}@{\extracolsep\fill}lcc}
    \toprule
    \multirow{2}{*}{Stage} & \multicolumn{2}{@{}c@{}}{Efficiency (\%)}\tabularnewline
    \cmidrule{2-3}
     & $\left|H\right\rangle $  & $\left|V\right\rangle $ \tabularnewline
    \toprule
    $\eta_{\mathrm{369}}$: \hyphenSI{369}{\nano\meter} coupling efficiency in QFC module& 70.8  & 70.8 \tabularnewline
    \midrule
    $T_{\mathrm{580}}$: optical path transmittance at \SI{580}{\nano\meter} of QFC module & 47.8 & 47.8\tabularnewline
    \midrule
    $T_{\mathrm{fib2}}$: coupling efficiency and transmittance of \hyphenSI{90}{\meter} SMF & 40.0 & 40.0\tabularnewline
    \midrule
    $\eta_{\mathrm{conv}}$: internal conversion efficiency  & 0.70 (\SI{2.00}{\watt}) & 0.75 (\SI{2.13}{\watt})\tabularnewline
    \midrule
    $\eta_{\mathrm{bw}}$: bandwidth matching efficiency between \hyphenSI{580}{\nano\meter} flying qubits and QM & 74.0 & 74.0\tabularnewline
    \midrule
    $\eta_\mathrm{storage}$: QM's device efficiency of \hyphenSI{1}{\micro\second} storage time & 19.5 & 18.3\tabularnewline
    \midrule
    $\eta_\mathrm{AOM}$: diffraction efficiency of AOM which is used for noise suppression & 80.0 & 80.0\tabularnewline
    \toprule
    $\eta$: overall efficiency of QFC module and QM node & 0.011 & 0.011\tabularnewline
    \bottomrule
    \end{tabular*}
    \label{tab2}
\end{table}

\backmatter




\bmhead{Acknowledgements}
This work was supported by the National Key Research and Development Program of China 
(No. 2024YFA1409403),
the National Natural Science Foundation of China 
(Nos. 11734015, 12204455, 12222411, 11821404, 12374336 and 12404572),
the Quantum Science and Technology-National Science and Technology Major Project 
(Nos. 2021ZD0301604 and 2021ZD0301200), 
the Strategic Priority Research Program of the Chinese Academy of Sciences 
(No. XDB1740000), 
and the Anhui Provincial Natural Science Foundation, China 
(No. 2508085MA005). 
Z.-Q.Z acknowledges the support from the Youth Innovation Promotion Association CAS.

\section*{Declarations}
\begin{itemize}

    \item Competing interests

    The authors declare no competing interests.

    \item Data availability

    The data that support the findings of the study are available from the corresponding author upon reasonable request.

    \item Materials availability

    The materials that support the findings of the study are available from the corresponding author upon reasonable request.

    \item Code availability

    The code used for data analysis and figure generation is available from the corresponding author upon reasonable request.

    \item Author contribution

    C.-X.W., Q.-Q.Y. constructed the trapped-ion node with the help of J.-M.C.
    Y.-Y.W. and T.-X.Z. constructed the quantum memory. Q.-Q.Y constructed the electronic control system for the ion trap with the help of J.-M.C.
    Q.-Q.Y and Y.-C.L constructed the QFC module with help from C.-X.W.
    P.-J.L. grew the $\rm ^{153}Eu^{3+}\colon\!Y_2SiO_5$ crystal with help from Z.-P.L.
    R.H., Y.-J.H.and Unitary Quantum Co., Ltd. provided technical support for the maintenance of the trapped-ion system.
    C.-X.W., J.-M.C., Y.-Y.W., T.-X.Z. and Z.-Q.Z. wrote the manuscript with the input from others.
    Z.-Q.Z. supervised the experiments on quantum memories and crystal growth. J.-M.C., Z.-Q.Z., Y.-F.H, and C.-F.L. designed the experiment and supervised the project. All authors discussed the experimental procedures and results.

\end{itemize}

\bibliography{ref}

\end{document}


\title[Article Title]{Supplementary:  Heterogeneous entanglement between a trapped ion and a solid-state quantum memory}

\author[1,2,3]{\fnm{Chen-Xu} \sur{Wang}}
\equalcont{These authors contributed equally to this work.}

\author[1,2,3]{\fnm{Yi-Yang} \sur{Wang}}
\equalcont{These authors contributed equally to this work.}

\author[1,2,3]{\fnm{Tian-Xiang} \sur{Zhu}}
\equalcont{These authors contributed equally to this work.}

\author[1,2,3]{\fnm{Qing-Quan} \sur{Yao}}
\equalcont{These authors contributed equally to this work.}

\author[1,2,3]{\fnm{Peng-Jun} \sur{Liang}}

\author[1,2,3]{\fnm{Yuan-Cong} \sur{Li}}

\author[1,2,3]{\fnm{Zi-Peng} \sur{Liu}}

\author[5]{\fnm{Ran} \sur{He}}

\author[1,2,3,4]{\fnm{Yong-Jian} \sur{Han}}

\author*[1,2,3,4]{\fnm{Jin-Ming} \sur{Cui}}\email{jmcui@ustc.edu.cn}

\author*[1,2,3,4]{\fnm{Zong-Quan} \sur{Zhou}}\email{zq\_zhou@ustc.edu.cn}

\author*[1,2,3,4]{\fnm{Yun-Feng} \sur{Huang}}\email{hyf@ustc.edu.cn}

\author*[1,2,3,4]{\fnm{Chuan-Feng} \sur{Li}}\email{cfli@ustc.edu.cn}

\author[1,2,3,4]{\fnm{Guang-Can} \sur{Guo}}

\affil[1]{\orgdiv{Laboratory of Quantum Information}, \orgname{University of Science and Technology of China}, \orgaddress{\city{Hefei}, \postcode{230026}, \country{China}}}

\affil[2]{\orgdiv{Anhui Province Key Laboratory of Quantum Network}, \orgname{University of Science and Technology of China}, \orgaddress{\city{Hefei}, \postcode{230026}, \country{China}}}

\affil[3]{\orgdiv{CAS Center For Excellence in Quantum Information and Quantum Physics}, \orgname{University of Science and Technology of China}, \orgaddress{\city{Hefei}, \postcode{230026}, \country{China}}}

\affil[4]{\orgdiv{Hefei National Laboratory}, \orgname{University of Science and Technology of China}, \orgaddress{\city{Hefei}, \postcode{230088}, \country{China}}}

\affil[5]{\orgname{Unitary Quantum Co., Ltd.}, \orgaddress{\city{Hefei}, \country{China}}}

\renewcommand{\thesection}{S\arabic{section}}
\renewcommand{\thefigure}{\arabic{figure}}
\captionsetup[figure]{name=Supplementary Fig.}
\setcounter{figure}{0}

\renewcommand{\thetable}{\arabic{table}}
\captionsetup[table]{name=Supplementary Table}
\setcounter{table}{0}
\renewcommand{\theequation}{S\arabic{equation}}
\setcounter{section}{0}
\setcounter{figure}{0}
\setcounter{table}{0}
\setcounter{equation}{0}
\maketitle

\section{The apparatus of the trapped-ion node}\label{apparatus}
The $^{171}\mathrm{Yb}^+$ ion is confined in a radio-frequency (RF) trap designed as a segmented-blade Paul trap, which is mounted inside a glass vacuum chamber using a ceramic holder, as shown in Supplementary Fig.~\ref{fig:TIsetup}a. 
The trap geometry provides an optical access with a numerical aperture (NA) of 0.64 along the direction of the applied magnetic field (defined as the $z$ axis). 
Laser beams for Doppler cooling, optical pumping, repumping, and pulsed excitation are delivered to the ion from the $x$-$y$ plane. 
A static magnetic field of approximately \SI{4}{G} is applied, resulting in a Zeeman splitting of \SI{11.22}{\mega\hertz} between the $\lvert 1 \rangle$ and $\lvert 1' \rangle$ states, as defined in the main text.

Along the $z$ axis, two objectives are arranged in a counter-propagating (top--bottom) geometry and share the same quantization axis defined by the magnetic field, as shown in Supplementary Fig.~\ref{fig:TIsetup}b. 
The upper objective, with an NA of 0.64 matching the optical access of the trap, is used to efficiently collect fluorescence photons that constitute the flying qubits. 
The lower objective, with an NA of 0.4, is dedicated to ion fluorescence imaging and qubit-state readout. 
Both objectives collect fluorescence from the same single ion.
As the high-NA (0.64) objective is aligned along the magnetic-field direction, it efficiently collects $\sigma_{+}$- and $\sigma_{-}$-polarized fluorescence photons. 
For high-fidelity ion-photon entanglement generation, it is essential to prevent population of the undesired excited-state Zeeman sublevels $^2P_{1/2}\lvert F=1, m=\pm1\rangle$, since spontaneous emission from these sublevels would degrade the correlation between the photon polarization and the states $|1'\rangle$ and $|1\rangle$. 
This suppression is achieved by employing a purely $\pi$-polarized excitation laser, defined with respect to the quantization axis set by the magnetic field, thereby 
selectively drives the transition from the $|0\rangle$ state to the excited $|e\rangle$ state.
To ensure the required polarization purity, a set of three-dimensional Helmholtz coils is used to finely adjust the orientation of the magnetic field, such that it is precisely aligned with the optical axis of the objectives. 
This alignment is crucial for achieving high-fidelity ion-photon entanglement in the present experiment.

\begin{figure*}[htbp]
    \centerline{\includegraphics[width=1\textwidth]{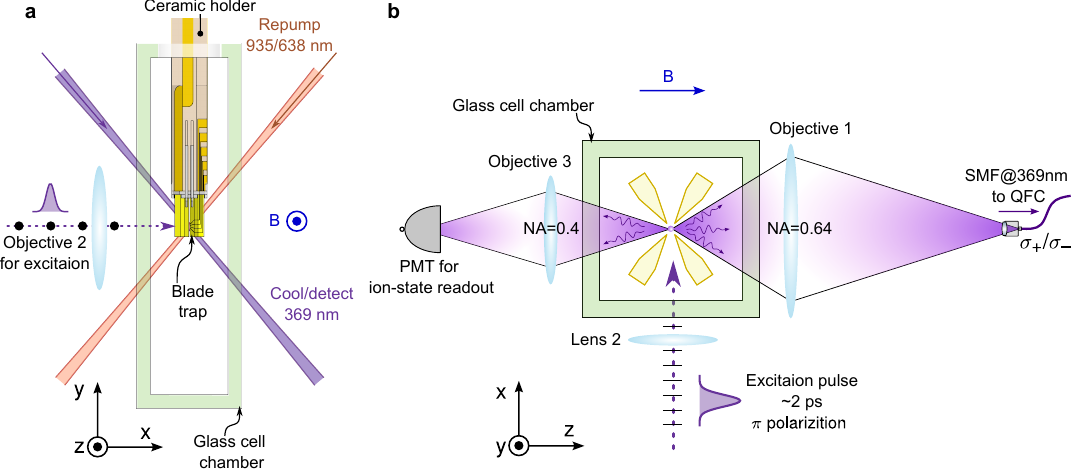}}
   \caption{
\textbf{Setup of the trapped-ion (TI) node.}
\textbf{a}, Lasers used for ion control. A 369-nm laser is employed for laser cooling, optical pumping, and state readout, while a 935-nm laser and a 638-nm laser are used for repumping.
\textbf{b}, Schematic of photon generation and ion-state readout. The single-pulse excitation laser is aligned to the ion along the $x$ direction with $\pi$ polarization. The magnetic field defining the quantization axis is applied along the $z$ direction. Two objectives arranged in a counter-propagating (top--bottom) geometry are used for fluorescence collection: Lens 1 (numerical aperture $(\mathrm{NA})=0.64$) collects photons for ion-photon entanglement generation and couples them into a single-mode fiber at \SI{369}{\nano\meter}, while Lens 3 ($\mathrm{NA}=0.4$) is dedicated to ion-state readout.
}
\label{fig:TIsetup}
\end{figure*}

\section{State preparation and measurement error of the TI node}\label{SPAM}
In this work, a single trapped $^{171}\mathrm{Yb}^+$ ion is employed as the trapped-ion (TI) node. During the entanglement generation process, the ion qubit is encoded in $\left|1\right\rangle$ and $\left|1'\right\rangle$ states, as described in the main text. For ion-state readout, the qubit is mapped onto the bright-dark basis ($\{\left|1\right\rangle, \left|0\right\rangle\}$) by using microwave (MW) sequence, as illustrated in Supplementary Fig~\ref{fig:SPAM}b.
Qubit initialization is realized via optical pumping using a resonant laser addressing the
$|\mathrm{S_{1/2}},F=1\rangle \leftrightarrow |\mathrm{P_{1/2}},F=0\rangle$ transition, supplemented by a sideband separated by \SI{2.105}{\giga\hertz} to compensate for population loss arising from the hyperfine splitting of the $\mathrm{P_{1/2}}$ manifold. During the pumping process, population in the $|\mathrm{S_{1/2}},F=1\rangle$ hyperfine manifold is excited to the $|\mathrm{P_{1/2}},F=1\rangle$ hyperfine manifold and subsequently decays via spontaneous emission to both the $|0\rangle$ state and $|\mathrm{S_{1/2}},F=1\rangle$ hyperfine manifold.
Owing to the \SI{12.642}{\giga\hertz} hyperfine splitting of the $\mathrm{S_{1/2}}$ manifold, the
$|0\rangle$ state is far detuned from the pumping laser and is therefore effectively forbidden from being re-excited. As a result, the optical pumping constitutes an effective cycle that terminates in the $|0\rangle$ state. After multiple pumping cycles, the ion is reliably initialized to the $|0\rangle$ state, as shown in Supplementary Fig~\ref{fig:SPAM}a.

State readout is performed via state-dependent resonant fluorescence, whereby only the bright state efficiently scatters photons. 
As shown in Supplementary Fig.~\ref{fig:SPAM}b, when the ion occupies the bright state, the state-readout laser resonantly drives the
$|\mathrm{S_{1/2}},F=1\rangle \rightarrow |e_1\rangle$ transition. 
After spontaneous emission from the excited state $|e_1\rangle$, the ion predominantly decays back to the $\mathrm{S_{1/2}}$ manifold. Population returning to the $|\mathrm{S_{1/2}},F=1\rangle$ hyperfine manifold can be re-excited by the state readout laser, while population decaying to the $|0\rangle$ state is far detuned and therefore forbidden from being re-excited. 
This mechanism establishes a closed fluorescence cycle for the bright state, enabling repeated photon scattering and efficient state readout. 
The emitted fluorescence photons are collected by the objective with a NA of 0.4, as shown in Supplementary Fig~\ref{fig:SPAM}b.
Fluorescence photons are recorded within a state-readout window of approximately \SI{180}{\micro\second}, enabling high-fidelity discrimination between the qubit states and yielding low overall state preparation and measurement (SPAM) error. 
A photon-count threshold of 1.5 is chosen to distinguish between the dark and bright states. The slightly reduced state-readout fidelity of the bright state arises from off-resonant excitation to the $|\mathrm{P_{1/2}},F=1\rangle$ manifold during the state-readout process, from which a fraction of the population decays to the $|0\rangle$ state, thereby leaving the fluorescence cycle and appearing dark. 
As a result, SPAM fidelities of 99.8\% for the dark state and 98.7\% for the bright state are obtained, as shown in Supplementary Fig.~\ref{fig:SPAM}c.

\begin{figure*}[htbp]
    \centerline{\includegraphics[width=1\textwidth]{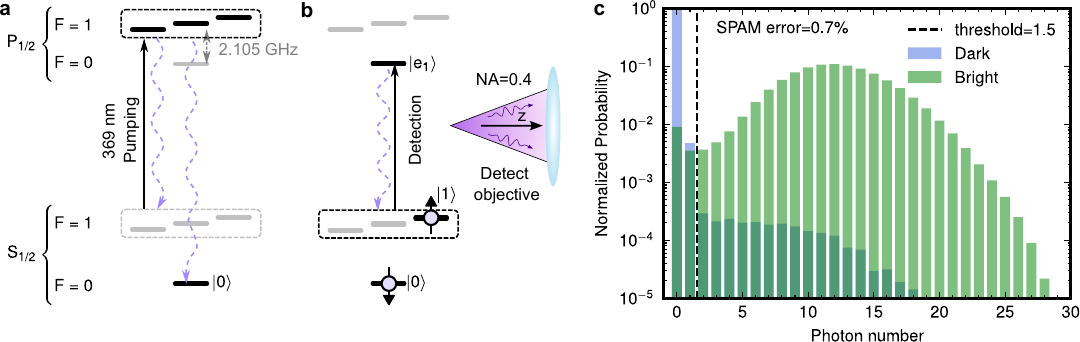}}
    \caption{
  \textbf{The state preparation and measurement (SPAM) details of the TI node.} \textbf{a}, The optical pumping. The ion qubit is pumped from the $|\mathrm{S_{1/2}},F=1\rangle$ hyperfine manifold to the $|\mathrm{P_{1/2}},F=1\rangle$ hyperfine manifold by a sideband at \SI{2.105}{\giga\hertz} of 369-nm laser, from there it can
decay to $|\mathrm{S_{1/2}}\rangle$ and can be reliably initialized to the $|0\rangle$ state after multiple pump cycles.
  \textbf{b}, The state readout of TI node. State readout is performed by resonant fluorescence. Bright-state ions are excited by the state readout laser from the $|\mathrm{S_{1/2}},F=1\rangle$ hyperfine manifold to the $|\mathrm{e_1}\rangle$ state, with scattered photons collected by a 0.4-NA objective.
  \textbf{c}, The initial states are prepared in the bright ($|1\rangle$) and dark ($|0\rangle$) states, respectively, over $10^6$ experimental shots.
  Each shot employs a state-readout time of \SI{180}{\micro\second}, yielding an average of 12.0 photons for the bright state.
  Using a photon-count threshold of 1.5, the measured state-readout fidelities are 99.8\% for the dark state and 98.7\% for the bright state,
corresponding to an overall SPAM error of 0.7\%.
}
    \label{fig:SPAM}
\end{figure*}
\section{Pulse excitation of the ion}\label{single_pulse}
To generate ion-photon entanglement, a short optical pulse is used to excite the ion from the $|0\rangle$ state to the $|e\rangle$ state. The detailed entanglement mechanism is described in the main text. The excitation pulses are derived from a pulse train produced by a mode-locked Ti:sapphire laser (Coherent Corporation, Mira-HP), emitting 739-nm pulses with a full width at half maximum (FWHM) of \SI{2.5}{\pico\second} at a repetition rate of \SI{76}{\mega\hertz}. The laser output is frequency-doubled using a $\mathrm{BiB_3O_6}$ (BIBO) crystal to generate 369-nm pulses at the same repetition rate, with a reduced FWHM of \SI{1.7}{\pico\second}.
To prevent the excitation from being influenced by neighboring pulses in the pulse train, a pulse-picker system is employed to select individual picosecond pulses.
The system is implemented using two acousto-optic modulators (AOMs) with a center frequency of \SI{400}{\mega\hertz}, controlled by a field-programmable gate array (FPGA), as shown in Supplementary Fig.~\ref{fig:pulse_excitation}a.
This configuration enables deterministic selection of single excitation pulses from the high-repetition-rate pulse train.

Excitation of the ion with the selected single pulses allows the excitation probability to be tuned by adjusting the pulse energy. To measure the excitation probability, the ion is first initialized to the $|0\rangle$ state and then excited by a single optical pulse. Following excitation to the $|e\rangle$ state, the ion decays via spontaneous emission to the $|0\rangle$, $|1\rangle$, and $|1'\rangle$ states. As a result, the ion occupies a bright state with a probability of $2/3$. The excitation probability is therefore inferred from the measured bright-state population as $P_\mathrm{e} = P_\mathrm{bright}/(2/3)$, where
$P_\mathrm{bright} = \frac{2A}{3}\sin^2\!\left(\frac{\alpha E^{\beta/2}}{2}\right),$ $A$, $\alpha$, and $\beta$ are fitting parameters and $E$ denotes the optical pulse energy~\cite{Feng2024}. As shown in Supplementary Fig.~\ref{fig:pulse_excitation}b, the measured excitation probability as a function of $\sqrt{E}$ exhibits excellent agreement with the theoretical model, confirming the effective operation of the single-pulse selection module. 
Ultimately, a single-pulse excitation probability of 96.0\% is achieved (see Supplementary Fig.~\ref{fig:pulse_excitation}b), including the SPAM error. Residual deviations of the fitted parameters from their ideal values are attributed to nonlinear response effects of the optical power meter.

The isolation performance of the dual-AOM scheme is independently characterized using a continuous-wave (CW) laser, yielding an extinction ratio exceeding $10^7{:}1$. This measurement reflects the intrinsic on-off contrast of the AOMs and serves as a technical characterization complementary to the single-pulse excitation tests. The single-pulse selection performance verified at the TI node is shown in Supplementary Fig.~\ref{fig:pulse_excitation}c. By employing a 30-ns photon-detection window starting from the onset of spontaneous emission, background counts are effectively suppressed, and a signal-to-noise ratio (SNR) exceeding $1800{:}1$ is achieved for the detection of 369-nm photons using photomultiplier tubes (PMTs).

\begin{figure*}[h]
    \centerline{\includegraphics[width=1\textwidth]{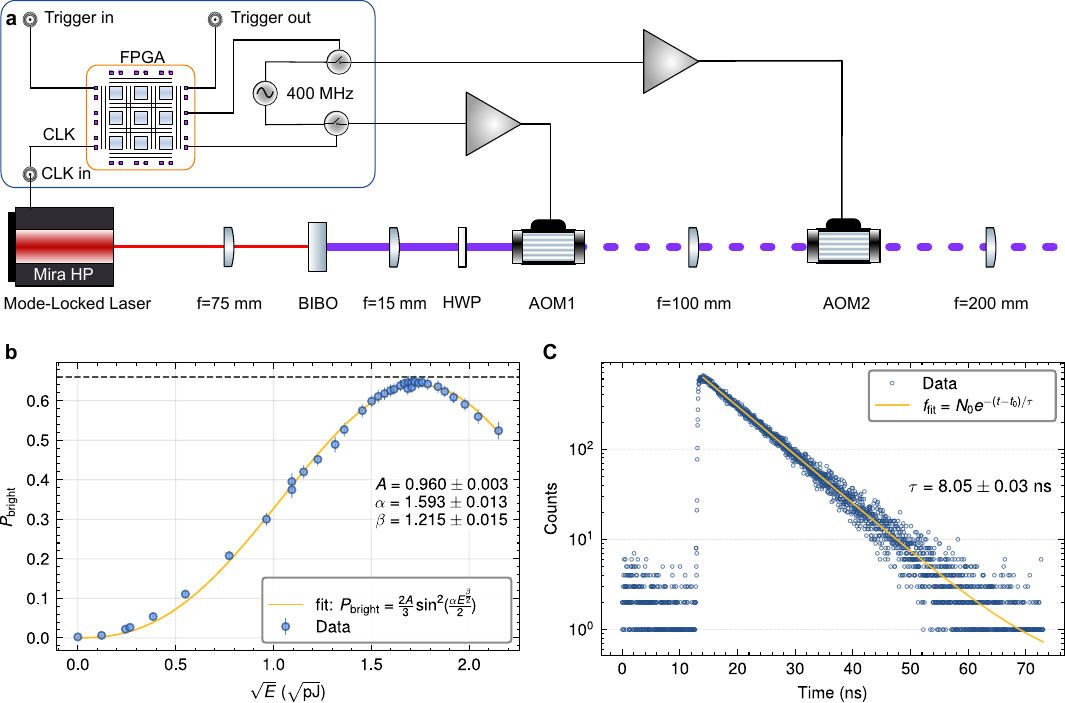}}
    \caption{\textbf{Single pulse generation and excitation for the trapped-ion node.} \textbf{a}, A train of pulses generated by a Mira-HP laser is passed through a $\mathrm{BiB_3O_6}$ (BIBO) crystal, frequency-doubled from \SI{739}{\nano\meter} to \SI{369}{\nano\meter}, and selected by two acousto-optic modulators (AOMs) with a central frequency of \SI{400}{\mega\hertz} for ion excitation. \textbf{b}, Probability of the ion being populated in the bright state after excitation as a function of the single-pulse energy. \textbf{c}, The time interval between the photon arrival time detected by the detector and the single-pulse excitation signal.}
    \label{fig:pulse_excitation}
\end{figure*}

\section{Photons detection and noise suppression}\label{Photonscandd} 

Photons retrieved from the solid-state quantum memory (QM) are detected using a superconducting nanowire single-photon detector (SNSPD) and precise photon-detection window of the detector is essential for enhancing the SNR. The arrival-time distribution of the retrieved photons is characterized using a time-to-digital converter (TDC), yielding a lifetime of $8.05 \pm 0.03$~\si{\nano\second} for the $|e\rangle$ state, as shown in Supplementary Fig.~\ref{fig:pulse_excitation}c. Accordingly, a photon-detection window of \SI{30}{\nano\second} is chosen, which captures approximately 97.6\% of the signal photons. The dominant noise sources include the dark counts of the SNSPD, environmental background noise, and noise generated in the QFC module. The noise suppression strategy described below primarily targets the latter contribution.

To eliminate temporally uncorrelated noise from the quantum frequency conversion (QFC) process, the QFC output is actively controlled using an AOM positioned before the 90-m single-mode fiber (SMF). When the AOM is deactivated, all noise contributions except for the intrinsic dark noise of the SNSPD and environment are effectively filtered out. When the AOM is activated, both signal photons and QFC-induced noise are allowed to propagate to the QM.
Given the finite absorption bandwidth of the QM (\SI{48.2}{\mega\hertz}), only noise photons spectrally overlapping with this bandwidth can be stored. Furthermore, following a total storage time of \SI{1}{\micro\second}, noise photons that are not temporally synchronized with the signal acquire a temporal offset relative to the retrieved signal photons. Consequently, only noise satisfying both the spectral acceptance of the QM and the \SI{30}{\nano\second} photon-detection window contributes to the measured noise, resulting in a measured SNR of 28{:}1.

\begin{figure*}[h]
    \centerline{\includegraphics[width=\textwidth]{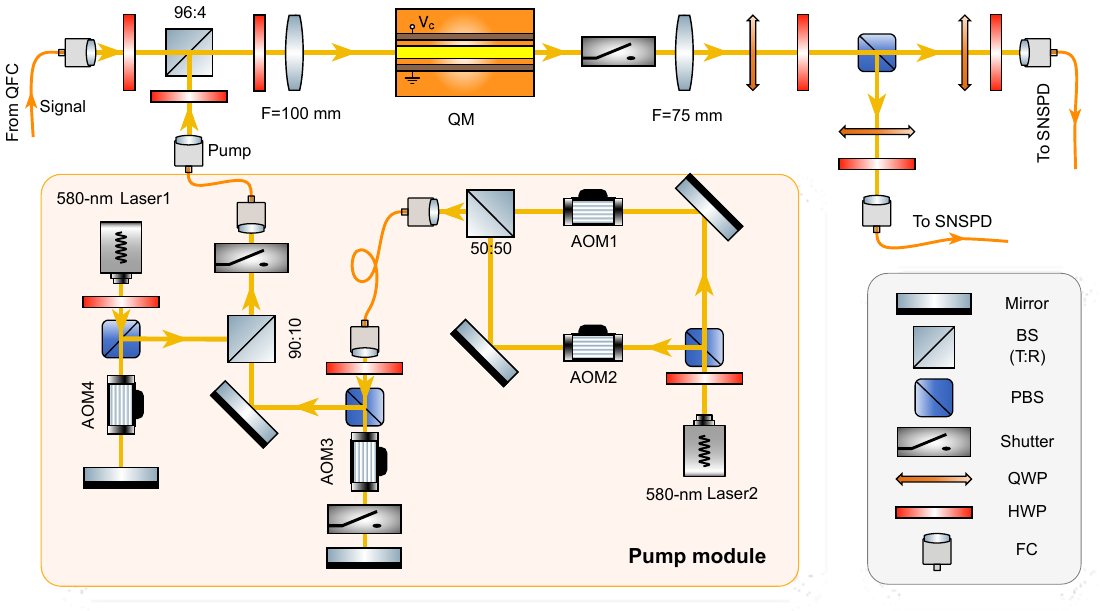}}
    \caption{\textbf{The experimental setup in the QM node.}
    The flying qubits from the QFC and the pump light are coupled into the QM via the transmission (96\%) and reflection (4\%) ports of the beam splitter (BS), respectively.
    The quantum state for the retrieved photon is then characterized using a polarization analysis module based on waveplates and PBS.
    The pump module consists of an absorption band preparation module and an atomic-frequency comb (AFC) preparation module, which are combined via a BS (90\%:10\%). The AFC preparation module comprises a double-pass AOM4 with a center frequency of +\SI{200}{\mega\hertz}. The absorption band preparation module employs cascaded AOMs (with center frequencies of +\SI{137}{\mega\hertz}, -\SI{137}{\mega\hertz}, and +\SI{200}{\mega\hertz} for AOM1, AOM2, and AOM3, respectively) to achieve a broader control bandwidth. Two mechanical shutters are employed to block the pump beam during single-photon storage process. Half-wave plate, HWP; quarter-wave plate, QWP; fiber collimator, FC.
    }
    \label{fig:QM node}
\end{figure*}
\section{QM preparation}\label{AFCpre}
\begin{figure*}[h]
    \centerline{\includegraphics[width=\textwidth]{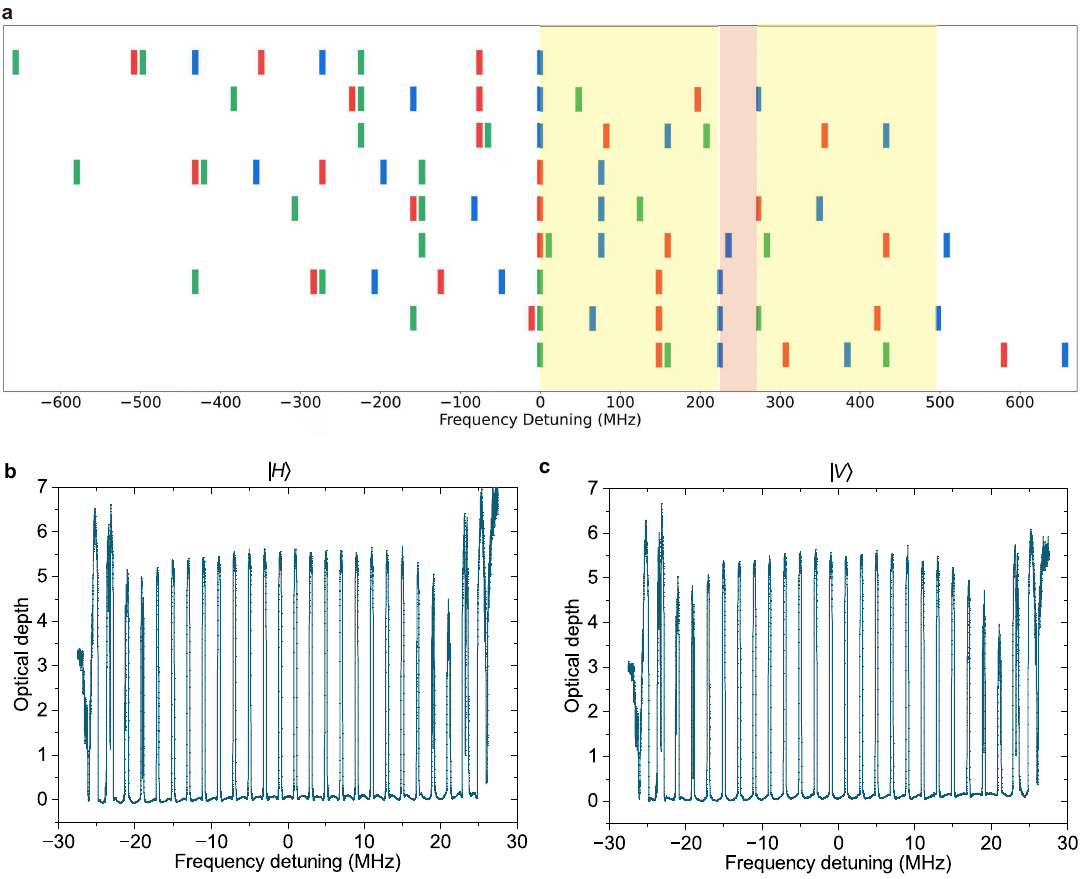}}
    \caption{\textbf{Preparation and measurement of AFC structures.} 
    \textbf{a}, Aligned energy-level diagram of 9 classes of site-2 $\rm ^{153}Eu^{3+}$ ions. From top to bottom, classes I to IX are shown. The yellow and orange areas are the range of the pump lights we applied. The blue, red, and green curves correspond to the transitions involving the ground states $\rm |1/2\rangle_g$, $\rm |3/2\rangle_g$, and $\rm |5/2\rangle_g$, respectively.
    \textbf{b--c}, The measured AFC structures when the input states are $|H\rangle$ (\textbf{b}) and $|V\rangle$ (\textbf{c}). }
    \label{fig:AFCpre}
\end{figure*}

The experimental setup in the QM node is presented in Supplementary Fig. ~\ref{fig:QM node}. For $\rm ^{153}Eu^{3+}\colon\!Y_2SiO_5$, the inhomogeneous broadening far exceeds the hyperfine splitting (Fig.~1i), leaving the individual hyperfine transitions unresolved. This leads to the observation of 9 distinct transitions upon light incidence, which correspond to the 9 classes of ions (Supplementary Fig.~\ref{fig:AFCpre}a).

Here we present a pumping strategy to enhance the effective absorption depth to achieve a higher storage efficiency. To enhance the effective absorption depth in the target band (the orange area of Supplementary Fig.~\ref{fig:AFCpre}a, with a center frequency of $f_0$ = \SI{248.6}{\mega\hertz}), a series of pulses is employed. First, a transmission band that includes the target region is successfully generated by employing a pump light centered at $f_0$. 
Next, by using two pump light in turn, which have the same chirping bandwidth of \SI{223.2}{\mega\hertz} and different center frequency of $f_0$+\SI{137}{\mega\hertz} and $f_0$-\SI{137}{\mega\hertz} (the yellow areas of Supplementary Fig.~\ref{fig:AFCpre}a), an enhanced absorption band with a bandwidth of \SI{48.2}{\mega\hertz} centered at $f_0$ is prepared. 

As an illustration, we can examine class IX ions (the bottom one in Supplementary Fig.~\ref{fig:AFCpre}a). For convenience, all transitions can be considered as a rightward broadening of \SI{497.2}{\mega\hertz}~\cite{zhu2022demand, Meng2026natpho}. 
The $\rm |1/2\rangle_g\rightarrow\rm |1/2\rangle_e$ transition is pumped in region [\SI{49.5}{\mega\hertz}, \SI{272.7}{\mega\hertz}].
The $\rm |1/2\rangle_g\rightarrow\rm |3/2\rangle_e$ transition is pumped in region [0, \SI{113.6}{\mega\hertz}].
The $\rm |3/2\rangle_g\rightarrow\rm |1/2\rangle_e$ transition is pumped in region [0, \SI{75.1}{\mega\hertz}] and [\SI{125.9}{\mega\hertz}, \SI{349.1}{\mega\hertz}].
The $\rm |3/2\rangle_g\rightarrow\rm |3/2\rangle_e$ transition is pumped in region [0, \SI{190}{\mega\hertz}].
The $\rm |5/2\rangle_g\rightarrow\rm |1/2\rangle_e$ transition is pumped in region [0, \SI{223.2}{\mega\hertz}] and [\SI{274}{\mega\hertz}, \SI{497.2}{\mega\hertz}].
The $\rm |5/2\rangle_g\rightarrow\rm |3/2\rangle_e$ transition is pumped in region [0, \SI{64.1}{\mega\hertz}] and [\SI{114.9}{\mega\hertz}, \SI{338.1}{\mega\hertz}].
The $\rm |5/2\rangle_g\rightarrow\rm |5/2\rangle_e$ transition is pumped in region [0, \SI{65.4}{\mega\hertz}].
In region [0, \SI{49.5}{\mega\hertz}], the population of $\rm |5/2\rangle_g$ is fully excited, so the $\rm |1/2\rangle_g\rightarrow\rm |1/2\rangle_e$ transition contribute to the target band.
And for region [\SI{75.1}{\mega\hertz}, \SI{125.9}{\mega\hertz}], the population of $\rm |5/2\rangle_g$ is partly excited, so the $\rm |3/2\rangle_g\rightarrow\rm |1/2\rangle_e$ transition contribute to the target band. 
Similar analysis could be applied to the remaining regions [\SI{64.1}{\mega\hertz}, \SI{113.6}{\mega\hertz}] and [\SI{223.2}{\mega\hertz}, \SI{272.7}{\mega\hertz}].
The final enhanced $d$ are $\sim$10.5 and $\sim$9 for $|H\rangle$ and $|V\rangle$, respectively, while the original $d$ for our sample is 5.24 ($|H\rangle$) and 4.66 ($|V\rangle$).

Then we use the parallel comb preparation scheme~\cite{jobez2016towards} to create an AFC structure with a comb spacing of \SI{2}{\mega\hertz} and a total bandwidth of \SI{48.2}{\mega\hertz}. 
The AFC structures are presented in Supplementary Fig~\ref{fig:AFCpre}a ($|H\rangle$) and b ($|V\rangle$).
The measured AFC efficiency is $43.3\pm0.2~\%$ ($39.8\pm0.2~\%$) for $|H\rangle$ ($|V\rangle$) at \SI{0.5}{\micro\second}. 

\begin{figure*}[h]
    \centerline{\includegraphics[width=1\textwidth]{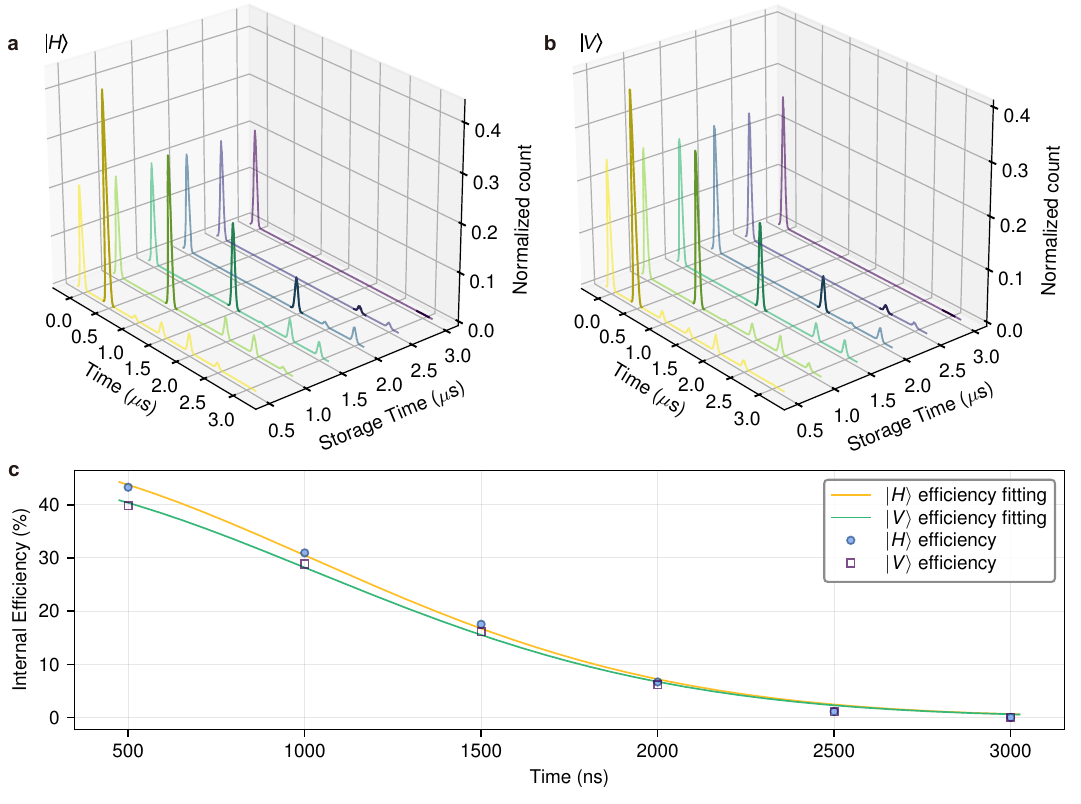}}
    \caption{\textbf{Storage efficiency of SMAFC memory.}
    \textbf{a--b}, The photon count histograms of SMAFC memories for $|H\rangle$ (\textbf{a}) and $|V\rangle$ (\textbf{b}), respectively.
    \textbf{c}, The internal efficiency of the memory for $|H\rangle$ (yellow) and $|V\rangle$ (green), as a function of storage time. According to the formula (\ref{f1}), the fitted $\gamma_\mathrm{comb}$ values are 259.8$\pm$5.7~\si{\kilo\hertz} ($|H\rangle$) and 259.0$\pm$6.0~\si{\kilo\hertz} ($|V\rangle$), respectively, which correspond to the $\mathcal{F}$ of 7.7$\pm$0.2 ($|H\rangle$) and 7.7$\pm$0.2 ($|V\rangle$). }
    \label{EXfig:SMAFC memory}
\end{figure*}
To achieve on-demand storage with electric control, the first electrical pulse with a voltage of \SI{8.6}{\volt} and a duration of \SI{100}{\nano\second} needs to be applied before the first echo appears, 
so that all of the echoes are suppressed until the second electrical pulse with reverse polarity is applied. If the second pulse is applied before the $n$'th order AFC echo located at $n\times\SI{0.5}{\micro\second}$, then the $n$'th AFC echo is readout in an on-demand fashion. 
The photon count histograms for Stark-modulated atomic frequency comb (SMAFC) storage of $\left|H\right\rangle$ and $\left|V\right\rangle$, are shown in Supplementary Fig~\ref{EXfig:SMAFC memory}a and b, respectively, with an input level of 0.12 photons per pulse. The internal storage efficiency of \SI{1}{\micro\second} SMAFC is $31.0\pm0.1\%$ ($28.9\pm0.1 \%$) for $|H\rangle$ ($|V\rangle$) as shown in Supplementary Fig~\ref{EXfig:SMAFC memory}c. Considering optical path loss, the device efficiency $\eta_\mathrm{storage}$ is $19.5\pm0.1 \%$ ($18.3\pm0.1 \%$).

The internal storage efficiency of SMAFC protocol is determined by absorption depth $d$ and the fineness of the AFC $\mathcal{F}$. If the combs are Gaussian, the internal storage efficiency of SMAFC protocol~\cite{horvath2021noise} is
\begin{equation}\label{f1}
\eta_\mathrm{storage}^\mathrm{internal}(t_\mathrm{storage})=B^{2}\dfrac{d^{2}}{\mathcal{F}^{2}}\exp\left(-B\dfrac{d}{\mathcal{F}}-2\pi B^{2}{t_\mathrm{storage}}^{2}\gamma_\mathrm{comb}^{2}\right),
\end{equation}
where $B=\sqrt{\pi}/\sqrt{4\ln2}$, $d$ is the absorption depth, $\gamma_\mathrm{comb}$ is the FWHM of the comb, $\Delta$ is the separation between the AFC peaks, $t_\mathrm{storage}=n/\Delta$ is the storage time, and $n$=1, 2, 3, ... is the order of the readout echo.

\section{Stability of QFC module and 90-m SMF}\label{QFCstab}
\begin{figure*}[htbp]
    \centerline{\includegraphics[width=1\textwidth]{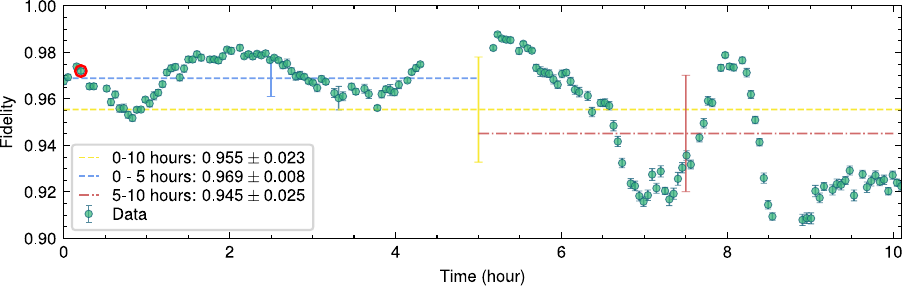}}
    \caption{\textbf{The stability of the process matrix fidelity of QFC and 90-m SMF.} The process matrix is measured by injecting high-power 369-nm reference laser. Over a 10-hour measurement period, the average process matrix fidelity reached $(95.5\pm2.3)$\% relative to the ideal identity matrix-with higher fidelity of $(96.9\pm0.8)$\% observed in the first 5 hours and a lower value of $(94.5\pm2.5)$\% in the subsequent 5 hours. The data points highlighted by the red circles correspond to the process matrix presented in Extended Fig.2.}
    \label{fig:QFC_stab}
\end{figure*}
The long-term stability of both the QFC module and the 90-m SMF is essential for our experiments. We first characterize the the 90-m SMF by injecting a set of polarization states into the 90-m SMF and reconstructing the corresponding process matrix at various times. The stability of the 90-m SMF is then quantified by evaluating the fidelity of the reconstructed process matrix with respect to the ideal identity process. The 90-m SMF is deployed outdoors with mechanical fixation and basic thermal protection.
The results show that the process matrix fidelity of the 90-m SMF remains above 99.0\% during the first 5~hours and above 98.5\% for up to 15~hours. 

Beyond the fiber-induced effects, the QFC module itself introduces additional instability. Excessive power of the 1018-nm pump laser, combined with fluctuations in the crystal temperature, leads to variations in both the waveguide coupling efficiency and the intrinsic difference-frequency generation (DFG) efficiency of the crystal. Temperature-induced variations in DFG efficiency constitute common-mode fluctuations for the two sub-loops of the QFC module and therefore do not affect the fidelity of the process matrix. In contrast, variations arising from changes in coupling efficiency are generally asynchronous between the two sub-loops, causing the overall process matrix of the QFC operation to deviate from the identity. Such deviations are non-unitary in nature and cannot be compensated by waveplates.
We characterize the long-term stability of the full 580-nm optical path, including the QFC module and the 90-m SMF. The corresponding fidelity stability was assessed via quantum process tomography over a 10-hour period (Supplementary Fig~\ref{fig:QFC_stab}). The average fidelity of the reconstructed process matrix is above 95\% throughout the measurement.

\section{Bandwidth matching between TI node and QM node}\label{bandwidth}
The spectral properties of the flying qubits emitted by the TI node are determined by the lifetime $\tau$ of the excited state $|e\rangle$, which is $\tau = \SI{8.12}{\nano\second}$. This lifetime corresponds to a natural linewidth of $\Gamma = 1/(2\pi\tau) = \SI{19.6}{\mega\hertz}$. Owing to Zeeman splitting, the two circularly polarized components of the emitted photons ($\sigma^+$ and $\sigma^-$) are separated in frequency by $\omega/2\pi = \SI{11.22}{\mega\hertz}$.
As a result, the spectrum of the flying qubits can be modeled as the incoherent sum of two Lorentzian line shapes with natural linewidth, each centered at frequencies $\Delta_f\pm \omega/4\pi$ after bandwidth matching, where $\Delta_f$ defined as the offset between the central frequency of the 580-nm photons and the QM. The normalized spectral density of the flying qubits is given by
\begin{equation}
    |S_p(f;\Delta_f)|^2=
    \frac{1}{2\pi \Gamma}
    \sum_{m=\pm1}
    \frac{(\Gamma/2)^2}
    {\left(f-f_0-\Delta_f+m\cdot\frac{\omega}{2}\cdot\frac{1}{2\pi}\right)^2+(\Gamma/2)^2}.
\end{equation}
This spectral structure leads to an intrinsic mismatch with the absorption bandwidth of the QM node, which is $B_\mathrm{QM}=\SI{48.2}{\mega\hertz}$. Even under perfect alignment of the central frequencies, the QM absorption spectrum cannot fully cover the spectral distribution of the flying qubits, resulting in a fundamentally limited bandwidth matching efficiency, as illustrated in Supplementary Fig.~\ref{fig:bandwidth_matching}a. 

\begin{figure*}[htbp]
    \centerline{\includegraphics[width=1\textwidth]{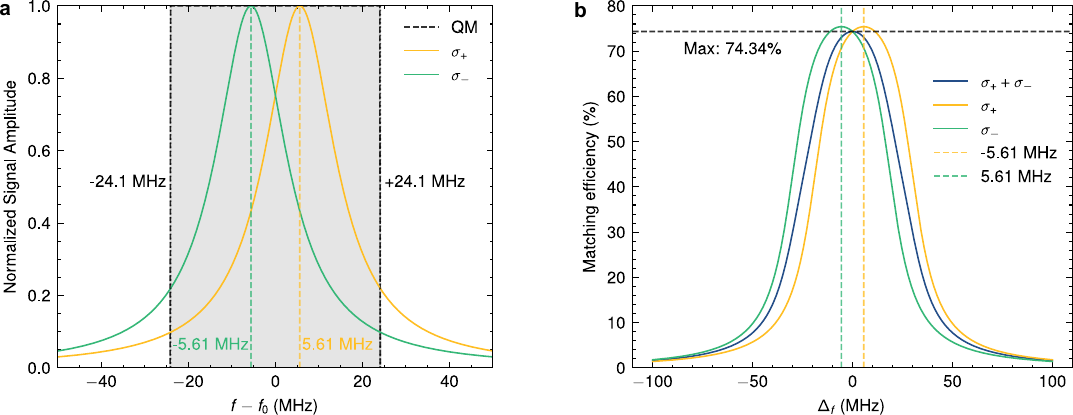}}
    \caption{\textbf{Bandwidth matching between TI node and QM node.} \textbf{a}, the theoretical maximum-normalized spectral density of 580-nm flying qubits (green and yellow line) at $\Delta_f=0$ and the storage bandwidth of QM (gray shaded region). $\Delta_f$ is defined as the offset between the central frequency of the 580-nm photons and the QM. \textbf{b}, The theoretical matching efficiency of bandwidth between TI node and QM node. The green (yellow) line denotes the bandwidth matching efficiency for $\sigma^+$ ($\sigma^-$) polarized photons and the QM. The blue line represents the total bandwidth matching efficiency.}
    \label{fig:bandwidth_matching}
\end{figure*}
In Supplementary Fig.~\ref{fig:bandwidth_matching}b, we present the bandwidth matching efficiency between the TI node and the QM node as a function of the center-frequency mismatch $\Delta_f$. The bandwidth matching efficiency is quantified by the normalized spectral overlap integral between the flying-qubit spectrum and the QM absorption profile  
\begin{equation}
\eta_\mathrm{bw}(\Delta_f)=
\int_{f_0-\frac{B_\mathrm{QM}}{2}}^{f_0+\frac{B_\mathrm{QM}}{2}}
|S_p(f;\Delta_f)|^2\,\mathrm{d}f ,
\end{equation}
Under ideal alignment ($\Delta_f = 0$), the maximum achievable bandwidth matching efficiency is theoretically limited to
\begin{equation}
\eta_\mathrm{bw}^\mathrm{theorymax} 
= 74.34\%.
\end{equation}
Experimentally, the spectral overlap (bandwidth matching efficiency) between the 580-nm photons and the Eu$^{3+}$ memory is measured to be approximately 74\%. This is determined by comparing photon counts when the QFC output is tuned into resonance with a prepared 48.2-MHz spectral pit versus when it is far detuned with Eu$^{3+}$ absorption by adjusting the wavelength of 1018-nm pump light.

\section{Error source analysis}\label{errsource}
\begin{table}[h!]
    \centering
    \caption{Error budget for entanglement infidelity.} 
    \begin{tabular}{lc} 
        \toprule %
        Error source\hspace{10em} & Infidelity\% \\ 
        \midrule %
        Ion decoherence &  $2.6\times10^{-6}$ \\ %
        Phase error due to photon's arrival-time jitter & $1.2\times10^{-4}$ \\ 
        SPAM error & 0.7 \\ 
        MW rotation  & 0.1 \\ 
        QFC & 3.1 \\  
        Pulse excitation  & 3.3 \\
        Collection of $\pi$ photons  & 0.5 \\ 
        Dark noise (SNSPD+environment)  & 2.7 \\ 
        Photon-state-detection & $2.9\times10^{-4}$ \\
        QM storage & 0.2 \\
        \midrule 
        Total error & 10.6 \\ 
        \bottomrule 
    \end{tabular}
    \label{tab3} 
\end{table} 
The main sources of error are summarized in Extended Table~\ref{tab3} and are discussed in detail below.
\begin{enumerate}
\item[(1)] Ion decoherence error. 
As the ion qubit is encoded in Zeeman-split energy levels, its transition frequency is inherently sensitive to ambient magnetic-field fluctuations. 
Between different experimental cycles, slow drifts of the magnetic field lead to run-to-run variations in the qubit resonance frequency. 
In contrast, the MW waveforms generated by the arbitrary waveform generator (AWG) are synthesized assuming a fixed reference frequency. 
As a result, an effective detuning arises in each experimental cycle, causing the ion-qubit state to accumulate an additional phase that varies from cycle to cycle.
This random phase evolution is subsequently mapped onto the relative phase $\phi$ of the ion-photon entangled state,
\begin{equation}
    |\Psi\rangle=\frac{1}{\sqrt{2}}\left(|1'\rangle|+\rangle+e^{-i\phi}|1\rangle|-\rangle\right),
\end{equation}
rendering $\phi$ a stochastic variable. 
Upon averaging over many experimental cycles, the resulting phase randomness leads to pure dephasing of the ion-photon entanglement.
The average fidelity limited by ion-qubit decoherence can be characterized as~\cite{Ball2016npj}
\begin{equation}
F_{\mathrm{avg}}(t) = \frac{1}{2}\left[1+\exp\!\big(-\chi(t)\big)\right],
\end{equation}
where $\chi(t)$ can be well approximated by a power-law form
\begin{equation}
\chi(t)=\left(\frac{t}{\tau_\mathrm{co}}\right)^a,
\end{equation}
with $t$ the evolution time, $\tau_\mathrm{co}$ the coherence time of the ion qubit, and the exponent $a=1,\dots,3$ depending on the noise spectrum~\cite{Degen2017RevModPhys}.
By assuming the noise source to be quasi-static, which leads to $a=2$, the average fidelity can be characterized as
\begin{equation}
F_{\mathrm{avg}}(t) = \frac{1}{2}\left[1+\exp\!\left(-\frac{t^2}{\tau_\mathrm{co}^2}\right)\right].
\end{equation}
Accordingly, the decoherence-induced error is given by $1 - F_{\mathrm{avg}}$. The $\tau_\mathrm{co}$ of the ion qubit is $(0.989\pm0.021)$~\si{\milli\second} probed with Ramsey interferometry (Supplementary Fig.~\ref{fig:Ramsey}). At the entanglement-state detection time, $t$ comprises a temporal delay of \SI{2.66}{\micro\second} (Methods) and a MW signal propagation time of \SI{0.51}{\micro\second}, giving rise to a decoherence error of $2.6\times10^{-6}$.
\begin{figure*}[htbp]
 \centerline{\includegraphics[width=1\textwidth]{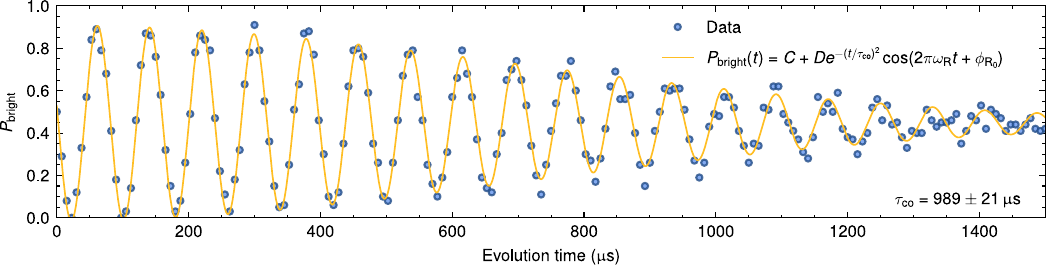}}
    \caption{\textbf{Decoherence of the ion qubit characterized by a Ramsey experiment.}
    The coherence time between the Zeeman sublevels $\lvert 1'\rangle$ and $\lvert 1\rangle$ is characterized using Ramsey interferometry and measured to be $(0.989 \pm 0.021)$~\si{\milli\second}. 
    The experimental data are fitted with
    $P_\mathrm{bright} = C + D e^{-(t/\tau_\mathrm{co})^2} \cos(2\pi \omega_\mathrm{R} t + \phi_\mathrm{R_0})$,
    where $C$ and $D$ are fitting constants, $\omega_\mathrm{R}$ is the Ramsey oscillation frequency, $\tau_\mathrm{co}$ denotes the coherence time, and $\phi_\mathrm{R_0}$ is the initial Ramsey phase.}
    \label{fig:Ramsey}
\end{figure*}

\item[(2)] Phase error due to photon-arrival-time jitter. 
As a consequence of synchronizing the ion-state readout to the photon detection (see in Supplementary information~\ref{IQM-tomo}), the timing jitter arising from the finite spontaneous-emission lifetime of the excited state is effectively removed from the entanglement measurement process. The dominant residual source of timing jitter during the quantum state tomography (QST) originates from the electronic signal transmission and processing chain, which introduces a device-dependent timing jitter. The device-dependent jitter in the photon arrival time leads to fluctuations in the relative phase accumulated between the two Zeeman sublevels $\left|1\right\rangle$ and $\left|1'\right\rangle$, which evolve at a frequency of $\omega=2\pi\times\SI{11.22}{\mega\hertz}$. By using the measurement circuits shown in Supplementary Fig.~\ref{fig:jitter}a--b, 
we determine a total timing jitter of 
$t_{\mathrm{total}}^{\mathrm{RMS}} = (0.310 \pm 0.012)~\si{\nano\second}$. 
This total jitter arises from independent contributions of 
$t_{\mathrm{AWG}}^{\mathrm{RMS}} = (0.305 \pm 0.012)~\si{\nano\second}$ from the arbitrary waveform generator (AWG) and 
$t_{\mathrm{trans}}^{\mathrm{RMS}} = (0.056 \pm 0.001)~\si{\nano\second}$ from the optical transceiver. 
Assuming statistical independence, the total RMS jitter is obtained by adding these contributions in quadrature,
$(t_{\mathrm{total}}^{\mathrm{RMS}})^2
=
(t_{\mathrm{AWG}}^{\mathrm{RMS}})^2
+
(t_{\mathrm{trans}}^{\mathrm{RMS}})^2,$
as shown in Supplementary Fig.~\ref{fig:jitter}c--d.
This temporal uncertainty corresponds to a phase uncertainty of
$\Delta\Phi = t_{\mathrm{total}}^{\mathrm{RMS}}\,\omega
= 2.19 \times 10^{-2}~\si{\radian}$.

Phase uncertainty suppresses the off-diagonal coherence of $|\Psi_\phi\rangle\langle\Psi_\phi|$ by averaging over the stochastic phase $\phi$, leading to a decay of the coherence factor $\langle e^{i\phi}\rangle$.
The resulting density matrix, expressed in the basis
$\{|1'\rangle|\sigma^+\rangle,\ |1\rangle|\sigma^-\rangle\}$,
can therefore be written by averaging $\phi$
\begin{equation}
\Big\langle |\Psi_\phi\rangle\langle\Psi_\phi|\Big\rangle_\phi 
=\frac{1}{2}
\begin{bmatrix}
1 & \langle e^{i\phi}\rangle \\
\langle e^{-i\phi}\rangle & 1
\end{bmatrix}.
\end{equation}
The fidelity with respect to the ideal target state $|\Psi\rangle=|\Psi_{\phi=0}\rangle$ is given by
\begin{equation}
F_\phi=\Big\langle \langle\Psi|\Psi_\phi\rangle\langle\Psi_\phi|\Psi\rangle\Big\rangle_\phi =\frac{1+|\langle e^{i\phi}\rangle|}{2}.
\end{equation}
In the small phase uncertainty ($\Delta\Phi \ll 1$) and quasi-static noise limit, the coherence factor can be approximated as
\begin{equation}
|\langle e^{i\phi}\rangle|\simeq
\exp\!\left(-\frac{\Delta\Phi^2}{2}\right)
\simeq1-\frac{\Delta\Phi^2}{2},
\end{equation}
which yields a phase-uncertainty-induced error
\begin{equation}
\epsilon_\phi \equiv 1-F_\phi\simeq\frac{\Delta\Phi^2}{4},
\end{equation}
which results in a phase-induced error of $1.2 \times 10^{-4}$~\cite{Degen2017RevModPhys,10.5555/AAI28196200}.

\begin{figure*}[htbp]
    \centerline{\includegraphics[width=1\textwidth]{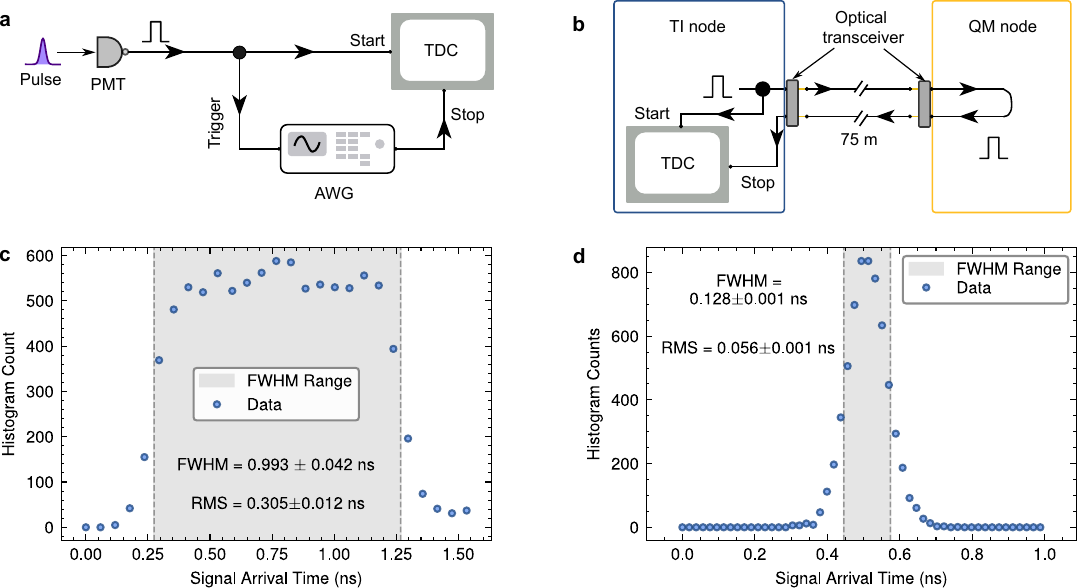}}
    \caption{\textbf{Timing jitter measurement of classical channels.} 
    \textbf{a}, The photon signals generated by the PMT in TI node are used as the start signals and the trigger of the AWG to produce electrical pulses that serve as the stop signals. The timing jitter of the AWG is quantified using a time-to-digital converter (TDC).
    \textbf{b}, To characterize the timing jitter of the optical transceiver, the electrical pulses are divided into two paths. One path serves as the start signals without passing through the classical channel, while the other acts as the stop signals after propagating through the classical channel.
    Timing jitters, characterized as root mean square (RMS) of time, are measured by TDC, showing timing jitter of $(0.305\pm0.012)$~\si{\nano\second} for the arbitrary waveform generator (AWG) due to its 1.25-GHz sampling rate (\textbf{c}) 
    and $(0.056\pm0.001)$~\si{\nano\second} for the optical transceiver (\textbf{d}).
    }
    \label{fig:jitter}
\end{figure*}

\item[(3)] SPAM error.  
The SPAM error in the ion-state preparation and readout, which employs a 0.4-NA objective, is measured to be 0.7\% (see Supplementary Information~\ref{SPAM}).

\item[(4)] MW rotation error.  
Imperfections in the MW-driven single-qubit rotations of ion contribute an error of 0.1\%, as characterized by randomized benchmarking measurements of the single-qubit MW gates.

\item[(5)] QFC infidelity.  
The infidelity of the QFC process including a 90-m single-mode fiber is obtained from the measured fidelity of the process matrix (Supplementary Information~\ref{QFCstab}), yielding a QFC infidelity of 3.1\%.

\item[(6)] Single-pulse excitation error.  
After accounting for the independently characterized SPAM error, the fidelity of the single-pulse excitation process is determined to be 96.7\% (Supplementary Information~\ref{single_pulse}), corresponding to a single-pulse excitation error of 3.3\%. SPAM and excitation errors are treated as independent.

\item[(7)] Collection error of $\pi$-polarized photons.  
Imaging aberrations in the imaging optical system lead to imperfect suppression of $\pi$-polarized photons in the single-mode fiber. By comparing the measured errors before and after the QFC stage under the same imaging optical system, this effect is estimated to contribute an error of 0.5\%.

\item[(8)] Dark noise error.  
Residual noise is dominated by SNSPD dark counts and background environmental light, defined as dark noise (see in Supplementary Information~\ref{Photonscandd}). Owing to the uniform statistics of dark noise across all measurement bases, its impact on the reconstructed density matrix can be expressed as
\begin{equation}
\rho = (1-p_\mathrm{noise})\rho_{\mathrm{ideal}} + p_\mathrm{noise}\frac{\mathbb{I}}{D},
\end{equation}
where $\rho_{\mathrm{ideal}}$ is a two-qubit maximally entangled state, $D=4$ is the Hilbert-space dimension, and $p_\mathrm{noise} = 1/(\mathrm{SNR}+1)$.  
The resulting infidelity is
\begin{equation}
\epsilon_{dark} = 1-((1-p_\mathrm{noise}) + \frac{p_\mathrm{noise}}{D}),
\end{equation}
yielding a dark-noise error of 2.7\%.

\item[(9)] Photon-state-detection error.  
Imperfect polarization discrimination by the PBS used for photon-state detection leads to errors in the measured photonic state. 
For a PBS with a finite extinction ratio, a fraction of the optical power leaks into the undesired polarization mode, characterized by a leakage probability
\begin{equation}
\label{eq:leakage}
\epsilon_\mathrm{PBS}=\frac{I_{\mathrm{leak}}}{I_{\mathrm{main}}},
\end{equation}
where $I_{\mathrm{main}}$ and $I_{\mathrm{leak}}$ denote the optical intensities in the desired and orthogonal polarization modes. 
This leakage effectively induces a polarization bit-flip error on the photonic qubit.
The effect of the PBS on the density matrix of TI-QM entanglement $|\Psi\rangle\langle\Psi|$ can therefore be modelled as a quantum channel acting on the photonic subsystem,
\begin{equation}
\mathcal{E}_{\mathrm{PBS}}(|\Psi\rangle\langle\Psi|)
=(1-\epsilon_\mathrm{PBS})|\Psi\rangle\langle\Psi|
+\epsilon_\mathrm{PBS}\,(\mathbb{I}\otimes X)|\Psi\rangle\langle\Psi|(\mathbb{I}\otimes X),
\end{equation}
where $X$ denotes the Pauli-$X$ operator in the polarization basis
$\{|\sigma^+\rangle,|\sigma^-\rangle\}$.
The infidelity with respect to the target state $|\Psi\rangle$ is therefore
\begin{equation}
1-F_\mathrm{PBS}=\langle\Psi|\mathcal{E}_{\mathrm{PBS}}(|\Psi\rangle\langle\Psi|)|\Psi\rangle=\epsilon_\mathrm{PBS}.
\end{equation}
The extinction ratios of the PBS used in the 580-nm photon detection is 3500 for both transmission and reflection, corresponding to a polarization-induced detection error of $2.9\times10^{-4}$.

\item[(10)] QM storage infidelity.  
The storage fidelities for the weak coherent input states $\left|H\right\rangle$, $\left|V\right\rangle$, $\left|H\right\rangle+\left|V\right\rangle$, and $\left|H\right\rangle+i\left|V\right\rangle$ are measured to be $99.97\pm0.01\%$, $99.90\pm0.01\%$, $99.81\pm0.01\%$, and $99.34\pm0.02\%$, respectively. Assuming that all input states occur with equal probability, this yields an average QM storage infidelity of $0.24\pm0.01\%$. 

All errors are treated as independent. This yields an estimated total error of 10.6\%, which is consistent with the experimentally measured fidelity of $(89.2 \pm 2.3)\%$.
\end{enumerate}

\section{Quantum state tomography}\label{IQM-tomo}
The TI-QM entanglement is characterized via QST, which relies on conditional joint measurements of the ion qubit and the retrieved photon polarization. The detection of 580-nm photons heralds the subsequent measurement of the ion state, enabling reconstruction of the two-qubit density matrix.
Specifically, the electrical output of the SNSPD is first sent to a FPGA. Then the FPGA generates a trigger signal for the AWG which controls the MW sequence for ion-state readout. 

For all measurement settings for the ion qubit, the sequence begins with a MW $\pi$ pulse that converts the $\left|1'\right\rangle$ state to the $\left|0\right\rangle$ state. An additional $\pi/2$ pulse with controlled phase is then optionally applied to choose the measurement basis. For the QM node, the photonic measurement basis is selected by adjusting the orientations of a set of waveplates using a motorized rotation stage, and the two polarization modes are detected by a two-channel SNSPD. 

For QST, the ion qubit is measured in three mutually unbiased bases, $\{|0\rangle, |1\rangle\}$, $\{(|0\rangle \pm |1\rangle)/\sqrt{2}\}$, and $\{(|0\rangle \pm i|1\rangle)/\sqrt{2}\}$, corresponding to measurements along the $Z$, $X$, and $Y$ axes, respectively. Similarly, the photonic qubit is measured in three polarization bases, $\{|H\rangle, |V\rangle\}$, $\{(|H\rangle \pm |V\rangle)/\sqrt{2}\}$, and $\{(|H\rangle \pm i|V\rangle)/\sqrt{2}\}$. The combination of these measurements yields $3\times3=9$ two-qubit product measurement settings, which form a tomographically complete set for reconstruction of the TI-QM density matrix. The two-qubit density matrix is reconstructed using maximum-likelihood estimation~\cite{Roos2004PRL,Haffner2005Nat} (MLE). 
The QST for TI-QM entanglement is identical to that used for ion-photon entanglement. The only difference is the photon-detection scheme, where a two-channel SNSPD is used for TI-QM entanglement, while two PMTs are used for ion-photon entanglement.

\section{\texorpdfstring{Phase evolution from the frequency splitting of the $\sigma^+$ and $\sigma^-$ photons}{Phase evolution from the frequency splitting of the sigma+ and sigma- photons}}\label{freq_diff}

The optical frequencies and detunings in this section are written as angular frequencies, and $\Gamma$ denotes the spontaneous decay rate, with $\Gamma/2\pi$ corresponding to the natural linewidth in hertz. We retain only the two collected decay channels that define the flying qubit,
$\ket{e}\rightarrow\ket{1'}+\sigma^+$ and
$\ket{e}\rightarrow\ket{1}+\sigma^-$.
The $\pi$-polarized channel is not part of the heralded qubit and is omitted from the following conditional two-channel description.
The relevant atomic levels are
\begin{equation}
    \ket{e}, \qquad \ket{1'}, \qquad \ket{1}.
    \label{Eq:freq_levels}
\end{equation}
Their energies are
\begin{equation}
    E_e = \hbar \omega_e, \qquad
    E_{1'} = \hbar \omega_{1'}, \qquad
    E_1 = \hbar \omega_1 .
    \label{Eq:atomic_energies_freq}
\end{equation}
The ground-state Zeeman splitting is
\begin{equation}
    \omega = \omega_1 - \omega_{1'} .
    \label{Eq:ground_splitting}
\end{equation}
The optical frequencies of the two emitted photons are
\begin{align}
    \nu_{+} &= \omega_e - \omega_{1'}, \\
    \nu_{-} &= \omega_e - \omega_1 .
    \label{Eq:optical_frequencies}
\end{align}
Therefore,
\begin{equation}
    \nu_{+}-\nu_{-}
    =
    \omega_1-\omega_{1'}
    =
    \omega ,
    \label{Eq:optical_frequency_difference}
\end{equation}
or equivalently
\begin{equation}
    \nu_{-}=\nu_{+}-\omega .
    \label{Eq:nu_minus}
\end{equation}
Thus the $\sigma^+$ photon associated with the decay to $\ket{1'}$ is centered at $\nu_{+}$, whereas the $\sigma^-$ photon associated with the decay to $\ket{1}$ is centered at $\nu_{-}$.

We take $t=0$ to be the time at which the excitation pulse prepares the atom in $\ket{e}$,
\begin{equation}
    \ket{\Psi(0)}=\ket{e}\ket{0_{\mathrm{ph}}}.
    \label{Eq:initial_excited_state}
\end{equation}
Within the two-channel model and assuming equal branching amplitudes into $\sigma^+$ and $\sigma^-$, spontaneous emission gives
\begin{equation}
    \begin{aligned}
        \ket{\Psi(t)}
        &=
        e^{-\Gamma t/2}\ket{e}\ket{0_{\mathrm{ph}}} \\
        &\quad
        +\sqrt{\frac{\Gamma}{2}}
        \int_0^t d\tau\,e^{-\Gamma \tau/2}
        \left[
        \ket{1'}\ket{\sigma^+,\tau;\nu_{+}}
        +
        \ket{1}\ket{\sigma^-,\tau;\nu_{-}}
        \right] .
    \end{aligned}
    \label{Eq:entangle_t}
\end{equation}
Here $\ket{\sigma^j,\tau;\nu_j}$ denotes a temporal photon mode emitted at time $\tau$, with polarization $\sigma^j$ and center frequency $\nu_j$ for $j\in\{+,-\}$.
The frequency-basis expansion of this temporal mode is
\begin{equation}
    \ket{\sigma^j,\tau;\nu_j}
    =
    \frac{1}{\sqrt{2\pi}}
    \int_{-\infty}^{+\infty}d\Delta\,
    e^{-i\Delta\tau}
    \ket{\sigma^j;\nu_j+\Delta},
    \label{Eq:temporal_frequency_basis}
\end{equation}
where $\Delta$ is the detuning from the corresponding center frequency.
After factoring out the equal branch amplitude $1/\sqrt{2}$ in Eq.~\ref{Eq:entangle_t}, the unnormalized spectral wave packet for either decay path can be written as
\begin{align}
    \ket{\phi_j(t)}
    &=
    \sqrt{\Gamma}
    \int_0^t d\tau\,e^{-\Gamma \tau/2}
    \frac{1}{\sqrt{2\pi}}
    \int_{-\infty}^{+\infty}d\Delta\,
    e^{-i\Delta\tau}
    \ket{\sigma^j;\nu_j+\Delta} \notag\\
    &=
    \sqrt{\frac{\Gamma}{2\pi}}
    \int_{-\infty}^{+\infty}d\Delta\,
    \left[
    \int_0^t d\tau\,
    e^{-(\Gamma/2+i\Delta)\tau}
    \right]
    \ket{\sigma^j;\nu_j+\Delta} \notag\\
    &=
    \int_{-\infty}^{+\infty}d\Delta\,
    F_{j,t}(\Delta)
    \ket{\sigma^j;\nu_j+\Delta},
    \label{Eq:single_photon_spectral_state}
\end{align}
with
\begin{equation}
    F_{j,t}(\Delta)
    =
    \sqrt{\frac{\Gamma}{2\pi}}
    \frac{1-e^{-(\Gamma/2+i\Delta)t}}
    {\Gamma/2+i\Delta}.
    \label{Eq:spectral_finite_t}
\end{equation}
For the two $\sigma$ transitions considered here, the natural linewidths are the same, so we write $F_{+,t}=F_{-,t}\equiv F_t$ when no confusion can arise.

The two terms in Eq.~\ref{Eq:entangle_t} are initially written in their own center-frequency frames. To compare their phase in a common optical frame, we rewrite the $\sigma^-$ component in the frame centered at $\nu_{+}$. Since
\begin{equation}
    \nu_{-}+\Delta
    =
    \nu_{+}+(\Delta-\omega),
    \label{Eq:common_frame_shift}
\end{equation}
the same $\sigma^-$ wave packet is
\begin{equation}
    \int d\Delta\,F_t(\Delta)\ket{\sigma^-;\nu_{-}+\Delta}
    =
    \int d\Delta\,F_t(\Delta+\omega)\ket{\sigma^-;\nu_{+}+\Delta}.
    \label{Eq:sigma_minus_common_frame}
\end{equation}
With the Fourier convention of Eq.~\ref{Eq:temporal_frequency_basis}, the shift
$F_t(\Delta+\omega)$ corresponds to a temporal phase factor,
\begin{equation}
    F_t(\Delta+\omega)
    \quad\Longleftrightarrow\quad
    f_t(\tau)e^{-i\omega\tau}.
    \label{Eq:fourier_shift}
\end{equation}
Therefore, in the common frame centered at $\nu_{+}$, Eq.~\ref{Eq:entangle_t} becomes
\begin{equation}
    \begin{aligned}
        \ket{\Psi(t)}
        &=
        e^{-\Gamma t/2}\ket{e}\ket{0_{\mathrm{ph}}} \\
        &\quad
        +\sqrt{\frac{\Gamma}{2}}
        \int_0^t d\tau\,e^{-\Gamma \tau/2}
        \left[
        \ket{1'}\ket{\sigma^+,\tau}_{\nu_{+}}
        +
        e^{-i\omega\tau}\ket{1}\ket{\sigma^-,\tau}_{\nu_{+}}
        \right] .
    \end{aligned}
    \label{Eq:entangle_common_frame}
\end{equation}
The frequency splitting has therefore been converted into a deterministic phase factor in the common optical frame.

For an ideal time-resolved herald corresponding to an emission time $t_0$, the no-emission term in Eq.~\ref{Eq:entangle_common_frame} is excluded and the integral is projected onto the temporal mode at $\tau=t_0$.
After normalization, the conditional ion-photon state is
\begin{equation}
    \ket{\Psi(t_0)}_{\mathrm{cond}}
    =
    \frac{1}{\sqrt{2}}
    \left[
    \ket{1'}\ket{\sigma^+,t_0}_{\nu_{+}}
    +
    e^{-i\omega t_0}
    \ket{1}\ket{\sigma^-,t_0}_{\nu_{+}}
    \right] .
    \label{Eq:conditional_emission_state}
\end{equation}
This equation makes two points explicit.
\begin{enumerate}
    \item Conditioned on the heralded $\sigma^\pm$ photon, the ion has already decayed from $\ket{e}$ to either $\ket{1'}$ or $\ket{1}$ and is entangled with the corresponding photon mode.
    \item With the time origin fixed by the excitation pulse, the relative phase in the common optical frame is determined by the spontaneous-emission time $t_0$.
\end{enumerate}
In the compact notation used in the main text, this phase is written as $\phi=\omega t_0+\phi_0$, where $\phi_0$ is a fixed offset set by the photon-mode phase convention and by the MW phase reference used for ion-state analysis. In the derivation below we set $\phi_0=0$, since it is a calibrated constant rather than a fluctuating phase.
There are two time coordinates in the discussion. First, in the excitation-pulse coordinate used above, $t=0$ is set by the picosecond excitation pulse. In this coordinate the emission time is $t_0$, and the recorded click time is
\begin{equation}
    t_c=t_0+T_{\mathrm{delay}},
    \label{Eq:click_time_delay}
\end{equation}
where $T_{\mathrm{delay}}$ is calibrated and includes optical propagation, QFC, QM storage and retrieval when present, and electronic delays. Second, for ion-state analysis, the same click both heralds the successful event and triggers the AWG waveform. It is therefore useful to define an AWG-triggered coordinate
\begin{equation}
    t_{\mathrm{A}}=t-t_c .
    \label{Eq:AWG_time_coordinate}
\end{equation}
In this coordinate the AWG starts at $t_{\mathrm{A}}=0$ with the programmed initial phase for the chosen analysis basis in every accepted trial. The same emission event is located at $t_{\mathrm{A}}=-T_{\mathrm{delay}}$, independent of the random value of $t_0$ in the excitation-pulse coordinate. Thus the random emission time changes the absolute trigger time $t_c$, but it does not make the initial phase of the triggered AWG waveform random. The remaining fixed phase relation between Eq.~\ref{Eq:conditional_emission_state} and the MW analysis basis is determined experimentally by the AWG phase calibration described in the main text.

We next check whether propagation adds another relative phase. Restoring the physical optical frequencies while keeping the frequency labels implicit in the photon-mode kets, a common propagation time $T$ gives
\begin{align}
    \ket{\sigma^+,t_0}
    &\rightarrow
    e^{-i\nu_{+}T}\ket{\sigma^+,t_0+T}, \\
    \ket{\sigma^-,t_0}
    &\rightarrow
    e^{-i\nu_{-}T}\ket{\sigma^-,t_0+T}.
    \label{Eq:photon_propagation}
\end{align}
The two atomic states simultaneously acquire
\begin{align}
    \ket{1'} &\rightarrow e^{-i\omega_{1'}T}\ket{1'}, \\
    \ket{1} &\rightarrow e^{-i\omega_1T}\ket{1}.
    \label{Eq:atomic_evolution}
\end{align}
The joint state after this time is therefore
\begin{align}
    \ket{\Psi(t_0,T)}
    =
    \frac{1}{\sqrt{2}}
    \Big[
    &
    e^{-i\omega_{1'}T}\ket{1'}
    \otimes
    e^{-i\nu_{+}T}\ket{\sigma^+,t_0+T}
    \notag\\
    +
    &
    e^{-i\omega t_0}
    e^{-i\omega_1T}\ket{1}
    \otimes
    e^{-i\nu_{-}T}\ket{\sigma^-,t_0+T}
    \Big] .
    \label{Eq:joint_after_propagation}
\end{align}
Using $\omega_{1'}+\nu_{+}=\omega_1+\nu_{-}=\omega_e$, this reduces to
\begin{equation}
    \ket{\Psi(t_0,T)}
    =
    \frac{1}{\sqrt{2}}e^{-i\omega_eT}
    \left[
    \ket{1'}\otimes\ket{\sigma^+,t_0+T}
    +
    e^{-i\omega t_0}
    \ket{1}\otimes\ket{\sigma^-,t_0+T}
    \right].
    \label{Eq:propagation_global_phase}
\end{equation}
Thus, after a common propagation delay $T$, the state has the same form as the compact expression used in the main text, apart from the global phase $e^{-i\omega_eT}$.

We now include the QM. This part acts on the spectral wave packet of each decay path, while the event-dependent factor $e^{-i\omega t_0}$ from Eq.~\ref{Eq:conditional_emission_state} is carried along. We do not assume that the QM is infinitesimally narrowband; instead, it is described by a frequency-dependent transfer function. For convenience, choose the frequency origin at the midpoint of the two optical frequencies,
\begin{equation}
    \nu_c
    =
    \frac{\nu_{+}+\nu_{-}}{2}
    =
    \nu_{+}-\frac{\omega}{2}
    =
    \nu_{-}+\frac{\omega}{2},
    \label{Eq:midpoint_frequency}
\end{equation}
and define
\begin{equation}
    \Omega=\nu-\nu_c .
    \label{Eq:Omega_midpoint}
\end{equation}
The two spectral wave packets can then be written in this common frequency frame as
\begin{align}
    \ket{\phi_+(t)}
    &=
    \int d\Omega\,
    F_t\!\left(\Omega-\frac{\omega}{2}\right)
    \ket{\sigma^+;\nu_c+\Omega}, \\
    \ket{\phi_-(t)}
    &=
    \int d\Omega\,
    F_t\!\left(\Omega+\frac{\omega}{2}\right)
    \ket{\sigma^-;\nu_c+\Omega}.
    \label{Eq:two_packets_midpoint_frame}
\end{align}
The offsets $\mp\omega/2$ simply express that the two photons are centered at different optical frequencies.

We model the QM by the mapping
\begin{equation}
    \ket{\sigma^\pm;\nu_c+\Omega}
    \longrightarrow
    G_\pm(\Omega,t_{\mathrm{storage}})\ket{M_\pm;\Omega},
    \label{Eq:QM_mapping}
\end{equation}
where $G_\pm(\Omega,t_{\mathrm{storage}})$ contains the spectral acceptance, phase response, and storage-time-dependent evolution of the memory for the two polarization modes. After successful storage for a time $t_{\mathrm{storage}}$, the two unnormalized memory wave packets are
\begin{align}
    \ket{\Phi_+(t_{\mathrm{storage}})}
    &=
    \int d\Omega\,
    G_+(\Omega,t_{\mathrm{storage}})
    F_t\!\left(\Omega-\frac{\omega}{2}\right)
    \ket{M_+;\Omega}, \\
    \ket{\Phi_-(t_{\mathrm{storage}})}
    &=
    \int d\Omega\,
    G_-(\Omega,t_{\mathrm{storage}})
    F_t\!\left(\Omega+\frac{\omega}{2}\right)
    \ket{M_-;\Omega}.
    \label{Eq:memory_wave_packets}
\end{align}
Their norms determine the storage probabilities,
\begin{equation}
    P_\pm
    =
    \braket{\Phi_\pm(t_{\mathrm{storage}})|\Phi_\pm(t_{\mathrm{storage}})} .
    \label{Eq:storage_probabilities}
\end{equation}
Conditioned on successful storage, the atom-memory state is
\begin{equation}
    \ket{\Psi_{\mathrm{QM}}(t_0,t_{\mathrm{storage}})}
    =
    \frac{
    \ket{1'}\ket{\Phi_+(t_{\mathrm{storage}})}
    +
    e^{-i\omega t_0}
    \ket{1}\ket{\Phi_-(t_{\mathrm{storage}})}
    }
    {\sqrt{P_+ + P_-}} .
    \label{Eq:general_atom_memory_state}
\end{equation}
Equation~\ref{Eq:general_atom_memory_state} is the general result: the QM does not erase the fact that the two decay paths have different optical center frequencies. It coherently selects and maps the corresponding spectral components into memory modes. A common transfer function changes the success probability and can add a fixed phase or reshape the accepted spectral wave packets. Unequal transfer functions $G_+\neq G_-$ can additionally change the relative amplitudes or make the two stored modes distinguishable.

As a transparent limit, consider a polarization-independent flat-top QM response centered at $\nu_c$,
\begin{equation}
    G_+(\Omega,t_{\mathrm{storage}})
    =
    G_-(\Omega,t_{\mathrm{storage}})
    =
    G(\Omega,t_{\mathrm{storage}}),
    \label{Eq:polarization_independent_QM}
\end{equation}
with
\begin{equation}
    G(\Omega,t_{\mathrm{storage}})
    =
    g_0 e^{-i\varphi_{\mathrm{QM}}(\Omega,t_{\mathrm{storage}})}
    W_{\mathrm{QM}}(\Omega),
    \label{Eq:QM_transfer_function}
\end{equation}
where $g_0$ is a frequency-independent storage amplitude that sets the overall storage probability and does not affect the relative phase, $\varphi_{\mathrm{QM}}(\Omega,t_{\mathrm{storage}})$ is the phase response of the QM, and
\begin{equation}
    W_{\mathrm{QM}}(\Omega)
    =
    \begin{cases}
    1, & |\Omega|\leq B_{\mathrm{QM}}/2,\\
    0, & |\Omega|>B_{\mathrm{QM}}/2 .
    \end{cases}
    \label{Eq:flat_top_window}
\end{equation}
In the experiment, the QM bandwidth is broad enough that the storage process only weakly modifies the entangled state and contributes, at most, a small fixed phase that is included in the phase calibration; to see analytically how the memory response can enter the relative phase, we now consider the simpler limiting case of a much narrower QM bandwidth.
If the accepted bandwidth is narrow enough that the spontaneous-emission spectrum is approximately constant over the accepted interval,
\begin{equation}
    F_t\!\left(\Omega-\frac{\omega}{2}\right)
    \simeq
    F_t\!\left(-\frac{\omega}{2}\right),
    \qquad
    F_t\!\left(\Omega+\frac{\omega}{2}\right)
    \simeq
    F_t\!\left(+\frac{\omega}{2}\right),
    \label{Eq:narrowband_approximation}
\end{equation}
then the two stored wave packets have the same spectral-temporal shape. Define the normalized memory modes
\begin{equation}
    \ket{M_\pm(t_{\mathrm{storage}})}
    =
    \mathcal{N}_M
    \int_{-B_{\mathrm{QM}}/2}^{B_{\mathrm{QM}}/2}
    d\Omega\,
    G(\Omega,t_{\mathrm{storage}})\ket{M_\pm;\Omega}.
    \label{Eq:normalized_memory_modes}
\end{equation}
The stored components become
\begin{align}
    \ket{\Phi_+(t_{\mathrm{storage}})}
    &\simeq
    F_t\!\left(-\frac{\omega}{2}\right)
    \ket{M_+(t_{\mathrm{storage}})}, \\
    \ket{\Phi_-(t_{\mathrm{storage}})}
    &\simeq
    F_t\!\left(+\frac{\omega}{2}\right)
    \ket{M_-(t_{\mathrm{storage}})} .
    \label{Eq:approx_memory_packets}
\end{align}
Because $|F_t(-\omega/2)|=|F_t(+\omega/2)|$, the normalized state reduces to
\begin{equation}
    \ket{\Psi_{\mathrm{QM}}(t_0,t_{\mathrm{storage}})}
    \simeq
    \frac{1}{\sqrt{2}}
    \left[
    \ket{1'}\ket{M_+(t_{\mathrm{storage}})}
    +
    e^{-i[\omega t_0+\phi_{\mathrm{filt}}]}
    \ket{1}\ket{M_-(t_{\mathrm{storage}})}
    \right],
    \label{Eq:atom_memory_narrowband_state}
\end{equation}
where the fixed filter-induced phase is
\begin{equation}
    \phi_{\mathrm{filt}}
    =
    \arg F_t\!\left(-\frac{\omega}{2}\right)
    -
    \arg F_t\!\left(+\frac{\omega}{2}\right).
    \label{Eq:filter_phase}
\end{equation}
In the long-time Lorentzian limit,
\begin{equation}
    F_t(\Delta)
    \rightarrow
    \sqrt{\frac{\Gamma}{2\pi}}
    \frac{1}{\Gamma/2+i\Delta},
    \label{Eq:long_time_lorentzian}
\end{equation}
one obtains
\begin{equation}
    \phi_{\mathrm{filt}}
    =
    2\tan^{-1}\!\left(\frac{\omega}{\Gamma}\right).
    \label{Eq:filter_phase_lorentzian}
\end{equation}

Therefore, for a polarization-independent QM, the storage time $t_{\mathrm{storage}}$ appears only in the common memory wave packet and does not introduce an additional relative phase. If the two memory modes acquire an extra relative phase during storage, for example due to a memory-mode splitting $\delta_{\mathrm{QM}}$, then the tracked phase should be replaced by
\begin{equation}
    \omega t_0+\phi_{\mathrm{filt}}
    \quad\longrightarrow\quad
    \omega t_0+\phi_{\mathrm{filt}}+\delta_{\mathrm{QM}}t_{\mathrm{storage}} .
    \label{Eq:storage_phase_replacement}
\end{equation}
Together with any constant calibrated offset $\phi_0$, this is the phase used to set the ion-analysis reference. In the absence of an untracked timing uncertainty or a polarization-dependent memory response, the frequency splitting of the $\sigma^+$ and $\sigma^-$ photons therefore produces a deterministic, trackable phase, not an intrinsic loss of entanglement fidelity.

\section{Uncertainty estimation under finite measurements}

In the QST, QPT, and CHSH measurements, only a finite number of experimental trials is performed for each measurement setting. For example, in QST, each product basis yields counts for four joint ion-photon outcomes; normalizing these counts by the total number of trials in that basis gives the corresponding projection probabilities. The density matrix and fidelity can then be reconstructed from the probability distributions measured over all bases, but this reconstruction alone does not provide statistical uncertainties. We therefore estimate the uncertainties of the QST and QPT fidelities, as well as of the CHSH parameter, by Monte Carlo resampling of the measured datasets, following a method commonly used in previous experiments~\cite{Huang2025Sciadv,DeGreve2013NC,Gao2012Nat}. For the QST of ion-photon and TI-QM entanglement, measurements are performed in nine product bases (three ion qubit bases $\times$ three photon polarization bases), with four outcomes for each basis, yielding 36 measured outcome counts. To generate one synthetic dataset, each measured count is used as the mean $\lambda$ of a Poisson distribution
\begin{equation}
    P(n)=e^{-\lambda}\frac{\lambda^n}{n!},
\end{equation}
from which a resampled count $n$ is drawn. We generate 1,000 such synthetic datasets. Each dataset is used to reconstruct a density matrix via MLE~\cite{Roos2004PRL,Haffner2005Nat}, from which the fidelity is computed. The reported uncertainty is the standard deviation of the fidelity distribution over all resampled datasets. The same resampling procedure is applied to estimate the uncertainties of the QPT fidelity and the CHSH parameter.

Alternatively, the Monte Carlo resampling can be performed by drawing the four outcome counts for each basis setting from a multinomial distribution~\cite{Krutyanskiy2023PRL, Pompili2021Sci}:
\begin{equation}
P_{\mathrm{multi}}(\mathbf{n}) =
\frac{{N}!}{n_1!\, n_2!\, n_3!\, n_4!}
\prod_{i=1}^{4} p_i^{n_i},
\end{equation}
where $N$ is the assigned total number of synthetic trials for that basis setting and $\mathbf{n}=(n_1,n_2,n_3,n_4)$ are the resampled outcome counts, with $\sum_i n_i=N$. The probability parameters are fixed by the measured data as $p_i=n_i^{\mathrm{obs}}/N_{\mathrm{obs}}$, where $n_i^{\mathrm{obs}}$ is the observed count of outcome $i$ and $N_{\mathrm{obs}}=\sum_{i=1}^{4} n_i^{\mathrm{obs}}$ is the total number of experimental trials in that basis.
For the two quantities reported in the main text, Supplementary Table~\ref{tab:poisson_vs_multinomial} compares the uncertainties obtained from Poisson and multinomial resampling. The two methods give comparable estimates; at the numerical precision used for reporting the main-text results, their difference is not resolved.
\begin{table}[htbp]
\centering
\caption{Comparison of Monte Carlo uncertainty estimation using Poisson and multinomial resampling (1,000 synthetic datasets for each method).}
\label{tab:poisson_vs_multinomial}
\begin{tabular}{@{}lccc@{}}
\toprule
\multirow{2}{*}{Items} & \multirow{2}{*}{Value} & \multicolumn{2}{c}{Uncertainty} \\
\cmidrule(lr){3-4}
 & & Poisson resampling & Multinomial resampling \\
\midrule
TI-QM entanglement fidelity & $89.2\%$ & $2.239\%$ & $2.234\%$ \\
CHSH parameter $S$ & $2.328$ & $0.0537$ & $0.0538$ \\
\bottomrule
\end{tabular}
\end{table}

\clearpage
\bibliography{ref}